\documentclass[10pt,letterpaper]{article}
\pdfoutput=1
\usepackage{jheppub}
\usepackage{dsfont}
\usepackage{amsfonts}
\usepackage{amsmath}
\allowdisplaybreaks[4]        
\usepackage{amssymb}
\usepackage{euscript}     
\usepackage{braket}
\usepackage{color}         
\usepackage{tensor}        
\usepackage{amsthm}
\usepackage{graphicx}
\usepackage{slashed}
\usepackage{subfigure}
\usepackage{bbm}
\usepackage[header,title,page,titletoc]{appendix}  

\usepackage[numbers]{natbib}  
\usepackage{float}

\newcommand{\cev}[1]{\reflectbox{\ensuremath{\vec{ \reflectbox{\ensuremath{#1}}}}}}

\newcommand{\bea}{\begin{eqnarray}}
\newcommand{\eea}{\end{eqnarray}}
\newcommand{\ba}{\begin{eqnarray}}
\newcommand{\ea}{\end{eqnarray}}

\newcommand{\beq}{\begin{equation}}
\newcommand{\eeq}{\end{equation} }
\newcommand{\beqa}{\begin{eqnarray}}
\newcommand{\eeqa}{\end{eqnarray}}
\newcommand{\beqar}{\begin{eqnarray*}}
\newcommand{\eeqar}{\end{eqnarray*}}

\newcommand{\mt}[1]{\textrm{\tiny #1}}

\newcommand{\crM}{\hat{a}^{\text{M}\dagger}_{k_z\vec{k}}}

\newcommand{\crR}{\hat{a}^{\text{R}\dagger}_{\omega\vec{k}}}
\newcommand{\aR}{\hat{a}^{\text{R}}_{\omega\vec{k}}}
\newcommand{\crRp}{\hat{a}^{R\dagger}_{\omega^\prime\vec{k}^\prime}}

\newcommand{\crL}{\hat{a}^{\text{L}\dagger}_{\omega\vec{k}}}
\newcommand{\aL}{\hat{a}^{\text{L}}_{\omega\vec{k}}}

\newcommand{\tM}{\text{M}}
\newcommand{\tR}{\text{R}}
\newcommand{\tL}{\text{L}}
\newcommand{\tU}{\text{U}}

\arxivnumber{arXiv:20nn.nnnnn}
\title{ \boldmath On operator growth and emergent Poincar\'e symmetries}

\author[a]{Javier M. Mag\'an}
\emailAdd{javier.magan@cab.cnea.gov.ar}
\author[b]{ and Joan Sim\'on}
\emailAdd{j.simon@ed.ac.uk}
\affiliation[a]{
Instituto Balseiro, Centro At\'omico Bariloche, 8400-S.C. de Bariloche, R\'io Negro, Argentina}
\affiliation[b]{
 School of Mathematics and Maxwell Institute for Mathematical Sciences,
University of Edinburgh, Edinburgh EH9 3FD, UK}

\date{\today}
\abstract{We consider operator growth for generic large-N gauge theories at finite temperature. Our analysis is performed in terms of Fourier modes, which do not mix with other operators as time evolves, and whose correlation functions are determined by their two-point functions alone, at leading order in the large-N limit. The algebra of these modes allows for a simple analysis of the operators with whom the initial operator mixes over time, and guarantees the existence of boundary CFT operators closing the bulk Poincar\'e algebra, describing the experience of infalling observers. We discuss several existing approaches to operator growth, such as number operators, proper energies, the many-body recursion method, quantum circuit complexity, and comment on its relation to classical chaos in black hole dynamics. The analysis evades the bulk vs boundary dichotomy and shows that all such approaches are the same at both sides of the holographic duality, a statement that simply rests on the equality between operator evolution itself. In the way, we show all these approaches have a natural formulation in terms of the Gelfand-Naimark-Segal (GNS) construction, which maps operator evolution to a more conventional quantum state evolution, and provides an extension of the notion of operator growth to QFT. 
}

\begin{document} 
\maketitle
\flushbottom

\newpage

\section{Introduction}
\label{sec:Introduction}

In the Schr\"odinger picture of quantum mechanics, unitary evolution mixes the initial state $\vert\psi  \rangle$
with other quantum states as time evolves
\begin{equation}\label{schr}
\vert\psi (t)\rangle = e^{-iHt}\,\vert\psi  \rangle =\sum_{n=0}^{\infty}\frac{(-iHt)^{n}}{n!}\,\vert\psi \rangle\equiv \sum_{n=0}^{\infty}\frac{(-it)^{n}}{n!}\,\vert\psi_{n} \rangle\;.
\end{equation}
Hence, solving for the time evolution amounts to understanding the states $\vert\psi_{n} \rangle \equiv H^{n}\vert\psi \rangle$, an understanding which is definitely challenging for chaotic Hamiltonians. Similarly, in the Heisenberg picture, unitary evolution mixes the initial operator $\mathcal{O}$ with other operators according to
\begin{equation}\label{Heis}
\mathcal{O}(t)=e^{iHt}\,\mathcal{O}\, e^{-iHt}=\sum\limits_{n=0}^{\infty}\frac{(it)^{n}}{n!}\,[H,\cdots,[H,\mathcal{O}]\cdots]\equiv \sum\limits_{n=0}^{\infty}\frac{(it)^{n}}{n!}\,\mathcal{O}_{n}\;.
\end{equation}
In this case, it is the understanding of the operators $\mathcal{O}_{n}  \equiv [H,\cdots,[H,\mathcal{O}]\cdots]$ that allows to solve for the time evolution, and ultimately determines any notion of \emph{operator growth} one might potentially define. 

The structure of Heisenberg's time evolution in chaotic systems has attracted some recent interest for several reasons. First, due to its expected connection to quantum chaos. It was found in 
\cite{Roberts:2018mnp, Qi:2018bje}, for the case of SYK \cite{Sachdev:1992fk,Kitaev}, that certain notion of operator size, to be reviewed in the main text, is related to out-of-time-ordered correlation (OTOC) functions \cite{1969JETP...28.1200L, Shenker:2014cwa,Maldacena:2015waa}. Second, because of the relation between operator growth, quantum complexity and the emergence of near horizon symmetries \cite{Susskind:2018tei,Magan:2018nmu,Lin:2019qwu,Barbon:2019tuq}. Finally, due to the broader connection between complexity and operator growth,
as discussed from different perspectives in \cite{Magan:2018nmu,Parker:2018yvk,Barbon:2019wsy,Bueno2019}, such as using Nielsen's geometric approach to quantum circuit complexity \cite{2005quant.ph..2070N,2007quant.ph..1004D} or the recursion method in many-body physics \cite{viswanath2008recursion}.\footnote{See \cite{Streicher:2019wek,Lucas:2019cxr,Lensky:2020ubw} for further work on operator growth in the context of SYK and \cite{Mousatov:2019xmc} for a more generic discussion in holographic theories.}

The main goal of this work is twofold. First and most important, to discuss the structure of the operators $\mathcal{O}_{n}$ in large-N theories, broadly understood as whenever large-N factorization holds \cite{tHooft:1973alw}, which include large-N holographic theories \cite{ElShowk:2011ag,Papadodimas:2012aq,Papadodimas:2013wnh,Papadodimas:2013jku,deBoer:2019kyr}. Second, to revisit some of the existent approaches to operator growth, apply them to large-N theories in the light of our previous analysis, and compare them with quantum circuit complexity and quantum chaos. Within the context of AdS/CFT \cite{Maldacena:1997re,Gubser:1998bc,Witten:1998qj}, the present approach, stemming on the analysis of the operators $\mathcal{O}_{n}$, evades the bulk vs boundary dichotomy, and makes manifest that \emph{any} notion of operator growth is the same at both sides of the duality, given the equivalence of Hilbert spaces, operator algebras, and Heisenberg time evolutions.

Albeit the first objective might seem a hopeless task, given the inherent complexity of a chaotic Hamiltonian, we will show how large-N factorization and generic finite temperature properties in relativistic QFTs 
completely determine the action of the operators $\mathcal{O}_{n}$ in most of the relevant states of the theory, at least if the Eigenstate Thermalization Hypothesis (ETH) \cite{PhysRevE.50.888} holds. Hence, we conclude that any notion of operator growth is determined by the behavior of the 2-pt functions alone, at leading order in the large-N limit.\footnote{As we will properly discuss in section~(\ref{sec:recursion}), this statement should not be associated with a similar statement in the context of the recursion method.}

As a byproduct of this discussion, we use the operators $\mathcal{O}_{n}$ in holographic theories to construct boundary CFT operators closing the \emph{bulk} Poincar\'e algebra.
Our construction is analogous to how the Poincar\'e algebra in free QFT is generated from the algebra of operators in the two Rindler wedges. It also uses the notion of mirror operators introduced in the context of holographic bulk reconstruction \cite{Papadodimas:2012aq,Papadodimas:2013wnh,Papadodimas:2013jku}. This emergent Poincar\'e algebra controls aspects of bulk infalling physics.

As for the second objective, we first note that quantum systems may have different notions of operator size depending on their nature and dynamics. However, all of them can be formulated as expectation values of simple operators within the Gelfand-Naimark-Segal (GNS) construction. The construction associates a Hilbert space to an algebra of operators, and maps Heisenberg's evolution to Schr\"odinger's evolution in the GNS Hilbert space. This allows extending notions of operator growth to QFT, and large-N theories in particular. As we will see, whenever large-N factorization holds, operator growth will be determined by 2-pt functions.

We then comment on natural size operators in these theories, mainly energy and number operators, stressing the exponential growth of the proper energy operator  \cite{Magan:2018nmu}. We also study the recursion method \cite{viswanath2008recursion}, typically used in many-body physics, in large-N theories.  We find a closed solution for the basis of orthonormalized operators that solves the 1d diffusion equation that appears in this approach. Finally, we describe the relation between Nielsen's quantum circuit complexity, operator growth, and chaos when one compares the complexity of formation between a pair of time evolved target states differing by an initial small perturbation.

This work is organized as follows. In section~\ref{sec:structure}, we discuss the operators $\mathcal{O}_{n}$ in large-N theories. In section~\ref{sec:algebra}, we first review the construction of the Poincar\'e algebra using the algebra of operators in the Rindler wedges in section~\ref{sec:rindler}. We then apply an analogous construction to holographic large-N theories in 
section~\ref{sec:GFFtoPoincare}. In section~\ref{sec:GNS}, we reformulate different notions of operator size in the literature using the GNS construction reviewed in appendix~\ref{App2}. We discuss natural notions of operator growth and the many-body recursion method in large-N theories in sections~\ref{sec:size-N} and \ref{sec:recursion}, respectively.
We close with a discussion connecting quantum circuit complexity and operator growth in section~\ref{sec:chaos}. A summary of our results and the logic purposed in this work are given in section~\ref{Discussion}.

\section{Operator evolution in large-N theories}
\label{sec:structure}

The goal of this section is to evaluate the series of nested commutators
\begin{equation}\label{lno}
\mathcal{O}_{n}\equiv [H,\cdots,[H,\mathcal{O}]\cdots]\;.
\end{equation}
controlling the Heisenberg time evolution \eqref{Heis} in generic large-N gauge theories for a subset of initial operators $\mathcal{O}$ where large-N factorization of correlation functions holds  \cite{tHooft:1973alw}.

Consider a local gauge-invariant scalar operator\footnote{The analysis of more generic smeared operators just follow from linearity as we comment further below.} $\mathcal{O}(t,\vec{x})$ in a gauge theory defined on $\mathbb{R}^{1,d-1}$. Its Fourier decomposition
\begin{equation}
   \mathcal{O}(t,\vec{x}) = \int_{\omega > 0} \frac{d\omega d^{d-1}\vec{k}}{(2\pi)^d} \left(\mathcal{O}_{\omega,\vec{k}}\,e^{-i\omega t + i\vec{k}\vec{x}} + \mathcal{O}^\dagger_{\omega,\vec{k}}\,e^{i\omega t - i\vec{k}\vec{x}}\right)\,,
\label{exparep}
\end{equation}
defines the non-local Fourier mode operators $\mathcal{O}_{\omega,\vec{k}}$ as
\begin{equation}
  \mathcal{O}_{\omega,\vec{k}}=\int dt\,d^{d-1}\vec{x} \,\mathcal{O}(t,\vec{x})\,e^{i\omega t-i\vec{k}\vec{x}} \,,
\end{equation}
with a similar expression for $\mathcal{O}^\dagger_{\omega,\vec{k}}$. 

Even though the energy $\omega$ and momentum $\vec{k}$ labels are \emph{not} related to each other by means of a dispersion relation, as it occurs for free quantum fields due to the classical equation of motion, the time evolution of $\mathcal{O}(t,\vec{x})$ remains trivial, as in the latter case, in this Fourier basis. Indeed, the nested commutators \eqref{lno} equal
\begin{equation}\label{no}
\mathcal{O}_{n}(t)= i^{-n}\frac{d^{n}}{dt^{n}}\mathcal{O}(t)=
\int\limits_{\omega > 0} \frac{d\omega d^{d-1}\vec{k}}{(2\pi)^d} \left((-\omega)^n\,\mathcal{O}_{\omega,\vec{k}}\,e^{-i\omega t + i\vec{k}\vec{x}} + \omega^n\,\mathcal{O}^\dagger_{\omega,\vec{k}}\,e^{i\omega t - i\vec{k}\vec{x}}\right)\,.
\end{equation}
In particular, each mode satisfies 
\begin{equation}
[H,\mathcal{O}_{\omega,\vec{k}}]=-\omega\,\mathcal{O}_{\omega,\vec{k}} \quad \Longrightarrow \quad  \mathcal{O}_{\omega,\vec{k}} (t)=e^{-i\omega t} \,\mathcal{O}_{\omega,\vec{k}}\,,
\end{equation}
and similarly for $\mathcal{O}_{\omega\vec{k}}^{\dagger}$. Hence, these Fourier mode operators do not mix with other operators as time evolves. For this reason, they are specially suited to study operator growth.

The Fourier decomposition \eqref{exparep} trades the problem of understanding operator growth, characterized by the operators $\mathcal{O}_{n}$, for the one of understanding $\mathcal{O}_{\omega,\vec{k}}$ and $\mathcal{O}_{\omega\vec{k}}^{\dagger}$. However, both these operators  are well understood in theories admitting a large-N expansion, within the regime where factorization of higher point functions holds. In particular, working at finite temperature, large-N factorization of thermal higher point correlation functions guarantees that finite temperature dynamics is determined by the 2-pt functions 
\begin{equation}\label{eq:exp}
\begin{aligned}
  Z^{-1}_\beta \text{Tr}\left(e^{-\beta H}\,\mathcal{O}_{\omega\vec{k}}\,\mathcal{O}_{\omega^\prime\vec{k}^\prime}\right) &= Z^{-1}_\beta \text{Tr}\left(e^{-\beta H}\,\mathcal{O}^\dagger_{\omega\vec{k}}\,\mathcal{O}^\dagger_{\omega^\prime\vec{k}^\prime}\right) = 0\,,\\
 Z^{-1}_\beta \text{Tr}\left(e^{-\beta H}\,\mathcal{O}_{\omega\vec{k}}\,\mathcal{O}^\dagger_{\omega^\prime\vec{k}^\prime}\right)&= G_\beta(\omega,\vec{k})\,\delta(\omega-\omega^\prime)\delta^{d-1}(\vec{k}-\vec{k}^\prime)\,,\\  
Z^{-1}_\beta \text{Tr}\left(e^{-\beta H}\, \mathcal{O}_{\omega\vec{k}}^{\dagger}\,\mathcal{O}_{\omega^\prime\vec{k}^\prime}\right) &=G_\beta(-\omega,-\vec{k})\,\delta(\omega-\omega^\prime)\delta^{d-1}(\vec{k}-\vec{k}^\prime)
\end{aligned}
\end{equation}
where $G_\beta(\omega,\vec{k})$ is the Fourier transform of the 2-pt function
\begin{equation}
 G_\beta(\omega,\vec{k}) \equiv \int dt\,d^{d-1}\vec{x}\, G_\beta(t,\vec{x})\,e^{i\omega t -i\vec{k}\vec{x}} \equiv\,Z_\beta^{-1}\,\int dt\,d^{d-1}\vec{x}\,\text{Tr}\left(e^{-\beta H}\,\mathcal{O}(t,\vec{x})\,\mathcal{O}(0,\vec{0})\right)\,e^{i\omega t -i\vec{k}\vec{x}}\,.
\end{equation}
It follows from \eqref{eq:exp} that the commutators of the Fourier mode operators are given by
\begin{equation}
\begin{aligned}
 Z^{-1}_\beta \text{Tr}\left(e^{-\beta H}\,\left[\mathcal{O}^\dagger_{\omega,\vec{k}}\,,\mathcal{O}^\dagger_{\omega^\prime,\vec{k}^\prime}\right]\right) &= Z^{-1}_\beta \text{Tr}\left(e^{-\beta H}\,\left[\mathcal{O}_{\omega,\vec{k}}\,,\mathcal{O}_{\omega^\prime,\vec{k}^\prime}\right]\right) = 0\,, \\
 Z^{-1}_\beta \text{Tr}\left(e^{-\beta H}\,\left[\mathcal{O}_{\omega,\vec{k}}\,,\mathcal{O}^\dagger_{\omega^\prime,\vec{k}^\prime}\right]\right) &= \left(G_\beta(\omega,\vec{k})-G_\beta(-\omega,-\vec{k})\right)\,\delta(\omega-\omega^\prime)\delta^{(d-1)}(\vec{k}-\vec{k}^\prime)\,,
\end{aligned}
\label{eq:N2p}
\end{equation}
up to $1/N$ corrections. These correlators assume that energy-momentum labels do not scale with $N$. Those high energy modes are not needed for the following reason. Local operators \eqref{exparep} have an infinite amount of energy, as it is typical in QFT. They contain modes up to infinite $\omega$ but they are not well defined operators. The actual operators we should consider are smeared versions of these 
\begin{equation}
\mathcal{O}=\int dt d^{d-1}\vec{x} f(x,t)\mathcal{O}(t,\vec{x}) = \int_{\omega > 0} \frac{d\omega d^{d-1}\vec{k}}{(2\pi)^d} \,\left(\tilde{f}(\omega,\vec{k})\,\mathcal{O}_{\omega,\vec{k}} + \tilde{f}^{*}(\omega,\vec{k})\,\mathcal{O}^\dagger_{\omega,\vec{k}}\right)\;,
\end{equation}
where
\begin{equation}
\tilde{f}(\omega,\vec{k})\equiv\int dt\,d^{d-1}\vec{x} \,f(t,\vec{x})\,e^{i\omega t-i\vec{k}\vec{x}} \,.
\end{equation}
To have a well defined (finite energy) operator $\mathcal{O}$ in the large-N limit we have to consider a smearing function $f(t,\vec{x})$ whose smoothness properties do not blow up in the limit. The Fourier transform of such function will exponentially suppress the modes $\mathcal{O}_{\omega,\vec{k}}$ with frequencies and wavelengths scaling with $N$ in the large-N limit. This smearing condition, together with large-N factorization, provides a good and precise definition of a ``simple'' operator in large-N QFT's.\footnote{In the context of spin systems, a ``simple'' operator is defined as one involving the product of an $\mathcal{O}(1)$ number of spins.}

Coming back to the algebra of the modes, as stressed in \cite{ElShowk:2011ag} and further developed in \cite{Papadodimas:2012aq}, due to large-N factorization, the same statements hold when inserting operators $\mathcal{P}_1$ and $\mathcal{P}_2$
\begin{equation}
\begin{aligned}
 Z^{-1}_\beta \text{Tr}\left(e^{-\beta H}\,\mathcal{P}_1\left[\mathcal{O}^\dagger_{\omega,\vec{k}}\,,\mathcal{O}^\dagger_{\omega^\prime,\vec{k}^\prime}\right] \mathcal{P}_2\right) &= Z^{-1}_\beta \text{Tr}\left(e^{-\beta H}\,\mathcal{P}_1 \left[\mathcal{O}_{\omega,\vec{k}}\,,\mathcal{O}_{\omega^\prime,\vec{k}^\prime}\right] \mathcal{P}_2\right) = 0\,, \\
 Z^{-1}_\beta \text{Tr}\left(e^{-\beta H}\,\mathcal{P}_1\left[\mathcal{O}_{\omega,\vec{k}}\,,\mathcal{O}^\dagger_{\omega^\prime,\vec{k}^\prime}\right] \mathcal{P}_2\right) &= \left(G_\beta(\omega,\vec{k})-G_\beta(-\omega,-\vec{k})\right)\,\delta(\omega-\omega^\prime)\delta^{(d-1)}(\vec{k}-\vec{k}^\prime)\cdot \\
 &\cdot \,\text{Tr}\left(e^{-\beta H}\,\mathcal{P}_1\mathcal{P}_2\right)
\label{eq:N2pa}
\end{aligned}
\end{equation}
involving a number of legs not scaling with $N$. Therefore, these commutators behave as c-numbers when inserted in correlation functions within this regime, a characteristic feature of free fields.

Correlators \eqref{eq:exp} and \eqref{eq:N2pa}, together with linearity, allow us to evaluate the expectation values of the nested commutators~(\ref{no}) in the thermal ensemble at any temperature. Furthermore, for the subset of large-N gauge theories satisfying the Eigenstate Thermalization Hypothesis (ETH) \cite{PhysRevE.50.888}, the set of states in the Hilbert space where the previous correlation functions hold is much larger since ETH ensures the same expectation values apply to most energy eigenstates compatible with the physical temperature. Therefore, in the large-N limit, the previous equations define the action of the operators $\mathcal{O}_{\omega,\vec{k}}$ and consequently of the $\mathcal{O}_{n}$ in the basis of eigenstates of the theory, up to a set of atypical energy eigenstates.\footnote{For a discussion on large-N factorization in chaotic theories and its relevance to ETH, see \cite{Papadodimas:2012aq,deBoer:2018ibj,deBoer:2019kyr}.} Notice that knowledge of this action and expectation values is enough to define the operators in this limit.\footnote{This is typical in probability theory. We can define a random variable by its associated probability distribution, or equivalently by giving all its moments. From a physical perspective, the second option is better since the moments are the ones being measured.}

Before ending this section, it is worth making a couple of closing remarks. First, the ``growth'' of the operator \eqref{exparep} in space, as defined by the nested commutators 
\begin{equation}
\begin{aligned}
  \left[P^j,\cdots,\left[P^j,\mathcal{O}(t,\vec{x})\right] \cdots\right] & = (-i)^{n}\frac{d^{n}}{dx_{j}^{n}} \mathcal{O}(t,\vec{x}) \\
&=  \int_{\omega > 0} \frac{d\omega d^{d-1}\vec{k}}{(2\pi)^d} \left[k_{j}^{n}\mathcal{O}_{\omega,\vec{k}}\,e^{-i\omega t + i\vec{k}\vec{x}} + (-k_{j})^{n}\mathcal{O}^\dagger_{\omega,\vec{k}}\,e^{i\omega t - i\vec{k}\vec{x}}\right]\,.
\end{aligned}
\label{pc}
\end{equation}
is also determined by the same 2-pt functions above. Second, in the large-$N$ limit we are working with, the generators of time $(H)$ and space $(\vec{P})$ translations effectively reduce to
\begin{equation}
  H=\int\limits_{\omega> 0} \frac{d\omega  d^{d-1}\vec{k}}{(2\pi)^d}\,\omega\,\mathcal{O}^{\dagger}_{\omega,\vec{k}}\mathcal{O}_{\omega,\vec{k}}\, \quad \text{and} \quad
  \vec{P}=\int\limits_{\omega> 0}  \frac{d\omega d^{d-1}\vec{k}}{(2\pi)^d}\, \vec{k}\,\mathcal{O}^{\dagger}_{\omega,\vec{k}}\mathcal{O}_{\omega,\vec{k}}\,.
\end{equation}

\paragraph{Modular time evolution.} Besides unitary time evolution, there is a second natural notion of evolution, modular time evolution, when restricting physics to subregions of spacetime (see \cite{Haag:1992hx} for a review). Modular time evolution is defined as the unitary evolution generated by the modular hamiltonian $H_{\textrm{mod}}$ in the region of interest
\begin{equation}
  \mathcal{O}(s)\equiv e^{isH_{\textrm{mod}}}\mathcal{O} e^{-isH_{\textrm{mod}}}=\rho^{-is}\mathcal{O}\rho^{is}\,,
\end{equation}
where $H_{\textrm{mod}}$ is related to the reduced density matrix\footnote{In QFT care has to be taken when defining these objects, but modular time evolution is well and unambiguously defined \cite{Haag:1992hx}. See \cite{Witten:2018lha} for a recent review.} in this region $\rho$, as $\rho=e^{-H_{\textrm{mod}}}$.

It is interesting to ask for the structure of operator evolution in this context.\footnote{Operator growth in the context of modular time evolution has also been considered recently in \cite{deBoer:2019uem}.} Since 
\begin{equation}
\mathcal{O}(s)=e^{isH_{\textrm{mod}}}\,\mathcal{O}\, e^{-isH_{\textrm{mod}}}=\sum\limits_{n=0}^{\infty}\frac{(is)^{n}}{n!}\,[H_{\textrm{mod}},\cdots,[H_{\textrm{mod}},\mathcal{O}]\cdots]\equiv \sum\limits_{n=0}^{\infty}\frac{(is)^{n}}{n!}\,\mathcal{O}^{\textrm{mod}}_{n}\;,
\end{equation}
this structure is determined by the operators
\begin{equation}
\mathcal{O}^{\textrm{mod}}_{n}\equiv [H_{\textrm{mod}},\cdots,[H_{\textrm{mod}},\mathcal{O}]\cdots]\;,
\end{equation}
which are the only ones with whom the initial operator mixes through modular time evolution. Proceeding as before, we can Fourier transform the fields, but against modular time evolution
\begin{equation}\label{modular}
\mathcal{O}_{\omega}=\int ds\, \,\mathcal{O}(s)\,e^{i\omega s}
\end{equation}
This allows for a simple computation of the action of $\mathcal{O}^{\textrm{mod}}_{n}$ in all eigenstates of the theory, in the vein of~(\ref{no}), if the correlators of the modular field modes are gaussian, as expected for holographic theories with free bulk duals.
%

\section{Emergent Poincar\'e algebra in holographic theories}
\label{sec:algebra}

Previous observations hold in large-N holographic theories. The dual bulk description of the CFT at finite temperature is a black hole \cite{Witten:1998zw} and the spectrum of low energy excitations is given by a small (not scaling with $N$) number of generalized free fields whose Fourier modes satisfy similar commutation relations to the ones in \eqref{eq:N2pa} (see below for a more precise account). Since the geometry near the black hole horizon is locally equivalent to a Rindler horizon, we can ask whether the exact relations between infalling and uniformly accelerated observers in free QFT in Rindler will continue to hold at the lowest order in a $1/N$ expansion in the holographic set-up.

In particular, free QFT in Rindler is Poincar\'e invariant. Motivated by the black hole scenario, we can ask how to build operators closing an exact Poincar\'e algebra out of the algebra of operators existing in the left and right Rindler wedges. We will review the construction of these operators in section~\ref{sec:rindler}. Since the latter only relies on the algebra satisfied by the Rindler creation/annihilation operators and this is the same algebra satisfied by the Fourier modes of the generalized free fields reconstructing the bulk excitations in the boundary CFT, it follows that applying the same Rindler construction to large-N holographic theories will give rise to boundary CFT operators closing the \emph{bulk} Poincar\'e algebra at lowest order in a $1/N$ expansion. The implications of this construction for operator growth will be discussed in the next section.

\subsection{From Rindler to Poincar\'e}
\label{sec:rindler}

Consider a free quantum scalar field of mass $m$ in $\mathbb{R}^{1,d}$. We want to review how to construct the relevant generators of the Poincar\'e algebra starting from the scalar description in both Rindler patches. To set some notation, let us decompose the space directions into $z$ and $\vec{x}$, with conjugate momentum $k_z$ and $\vec{k}$, respectively, so that $z$ corresponds to the direction along which the Rindler observer is uniformly accelerated. Local fields in the right Rindler patch can be expanded in terms of creation and annihilation operators
\begin{equation}
  \aR\,\,\,\,,\,\,\, \crR 
\end{equation}
satisfying the standard commutation relations 
\begin{equation}
  \left[\aR,\,\crRp\right] = \delta(\omega-\omega^\prime)\delta^{(d-1)}(\vec{k}-\vec{k}^\prime)\,.
\label{eq:right-rindler}  
\end{equation}
Since $\omega$ stands for the Rindler energy, time evolution in this Rindler wedge is generated by the operator
\begin{equation}
  \hat{H}^{\textrm{R}}=\int\limits d\omega\,d\vec{k}\, \omega\, \crR\aR\,.
\end{equation}
It follows that each operator $\aR$ evolves as
\begin{equation}
[\hat{H}^{\textrm{R}},\aR]= -\omega\,\aR \quad \Rightarrow \quad \aR (t)= e^{-i\omega t}\,\aR
\end{equation}
as usual in free quantum field theory, with a similar expression for $\crR (t)$. Linearity extends these claims to local quantum fields. Notice Rindler time is labelled as $t$. Minkowski operators and coordinates will carry an $M$ superscript.

Consider a thermal state $\rho_{\beta}$ in the right Rindler wedge quantum field theory (QFT) with $\beta=\frac{2\pi}{a}$, where $a$ stands for the proper acceleration defining the Rindler frame. It is well known the latter is the reduced density matrix of the pure state $\vert 0_{\text{M}}\rangle$, i.e. the vacuum of the QFT in Minkowski (see \cite{Crispino:2007eb} for a review and more details about QFT in Rindler space). This purification involves duplication of the operator algebra giving rise to a new set of operators
\begin{equation}
\aL\,\,\,\,,\,\,\, \crL 
\label{eq:left-rindler}
\end{equation}
which correspond to the creation and annihilation operators in the left Rindler wedge from the perspective of the full Minkowski spacetime. These operators commute with the original ones \eqref{eq:right-rindler} and altogether form a complete basis of operators in the Minkowski Hilbert space. This matches the general Gelfand-Naimark-Segal (GNS) representation of thermal states reviewed in appendix~\ref{App2}, for later convenience. In this appendix, the origin of the duplication of the algebra responsible for the canonical purification of the thermal Rindler state $\rho_\beta$ in the current discussion is explained.

The Minkowski vacuum $|0_\tM\rangle$ is determined by the relations 
\begin{equation}
  \left(\hat{a}^\tL_{\omega\vec{k}} - e^{-\pi\frac{\omega}{a}}\,\hat{a}^{\tR\dagger}_{\omega\,(-\vec{k})}\right)|0_\tM\rangle =  \left(\hat{a}^\tR_{\omega\vec{k}} - e^{-\pi\frac{\omega}{a}}\,\hat{a}^{\tL\dagger}_{\omega\,(-\vec{k})}\right) |0_\tM\rangle = 0\,.
\label{trind}  
\end{equation}
These allow to write the action of both $\aL$ and $\crL$ on $|0_\tM\rangle$ in terms of operators acting on the right wedge. Notice that in our conventions, those of Ref. \cite{Crispino:2007eb}, Rindler time in the left wedge also runs in the same direction as in the right Rindler wedge. To get the opposite conventions typically used in black hole physics and holography \cite{Maldacena:2001kr}, one needs to perform the replacement $\hat{a}^{\tL\dagger}_{\omega\,(-\vec{k})}\to \hat{a}^{\tL\dagger}_{\omega\,\vec{k}}$. 

Given this complete basis of operators, there are two different bases one can introduce which will be relevant in what follows. The first is the set of creation and annihilation operators associated with Minkowski time evolution 
\begin{equation}
  \hat{a}^\tM_{k_z\vec{k}} = \int^\infty_0 \frac{d\omega}{\sqrt{2\pi a\,\omega_{\vec{k}}}}\frac{1}{\sqrt{1-e^{-2\pi\omega/a}}} \left[e^{i \vartheta(k_z) \frac{\omega}{a}}
  \left(\hat{a}^\tL_{\omega\vec{k}} - e^{-\pi\frac{\omega}{a}}\,\hat{a}^{\tR\dagger}_{\omega\,(-\vec{k})}\right) + e^{-i\vartheta(k_z)\frac{\omega}{a}} \left(\hat{a}^\tR_{\omega\vec{k}} - e^{-\pi\frac{\omega}{a}}\,\hat{a}^{\tL\dagger}_{\omega\,(-\vec{k})}\right)\right]\,,
\label{eq:mink-op}
\end{equation}
where 
\begin{equation}
  \vartheta (k_z) = \frac{1}{2}\log \left(\frac{\omega_{\vec{k}}+k_z}{\omega_{\vec{k}}-k_z}\right)\,,
\end{equation}
is the standard rapidity in relativistic physics and $\omega_{\vec{k}}^{2}=m^{2}+k_{z}^{2}+\vert\vec{k}\vert^{2}$ is the on-shell Minkowski frequency carried by each mode. In this basis, the generators of Minkoswki time and spatial $z$ translations are the standard expressions
\begin{equation}
\begin{aligned}
  \hat{H}^{\tM}&=\int\limits dk_{z}\,  d^{d-1}\vec{k}\,\omega_{\vec{k}}\,\hat{a}^{\tM\dagger}_{k_z\vec{k}}\hat{a}^\tM_{k_z\vec{k}}\,,\\
  \hat{P}_{z}^{\tM}&=\int\limits  dk_{z}\,d^{d-1}\vec{k}\,k_{z}\,\hat{a}^{\tM\dagger}_{k_z\vec{k}}\hat{a}^\tM_{k_z\vec{k}}\;,
\end{aligned}
\label{eq:Mink-gen}
\end{equation}
whereas the number operator equals
\begin{equation}
\hat{N}^{\tM}=\int\limits  dk_{z}\,d^{d-1}\vec{k}\,\hat{a}^{\tM\dagger}_{k_z\vec{k}}\hat{a}^\tM_{k_z\vec{k}}\;.
\end{equation}
Notice that using \eqref{trind}, any Minkowski mode can be easily generated by acting with operators in the right wedge
\begin{equation}\label{ami3}
\crM \vert 0_{M}\rangle = \int^\infty_0 \frac{d\omega}{\sqrt{2\pi a\,\omega_{\vec{k}}}} \sqrt{2\sinh \frac{\pi\omega}{a}} \left[e^{-i \frac{\omega}{a}\left(\vartheta(k_z) + i\frac{\pi}{2}\right)}  \,\hat{a}^{R}_{\omega\,(-\vec{k})} + e^{i \frac{\omega}{a}\left(\vartheta(k_z) + i\frac{\pi}{2}\right)} \hat{a}^{R\dagger}_{\omega\vec{k}} \right]   \vert 0_{\text{M}}\rangle\,.
\end{equation}
The second basis is the Unruh basis defined by appropriate normalisation of the operators annihilating $|0_\text{M}\rangle$ in
\eqref{trind}
\begin{equation}\label{eq:unruh1}
b_{+\omega,\vec{k}}= \frac{1}{\sqrt{1-e^{-2\pi\frac{\omega}{a}}}}\left( \hat{a}^{\tL}_{\omega\vec{k}} - e^{-\pi\frac{\omega}{a}}\,\hat{a}^{\tR\dagger}_{\omega\,(-\vec{k})}\right) \,, \quad
b_{-\omega,\vec{k}}=\frac{1}{\sqrt{1-e^{-2\pi\frac{\omega}{a}}}}\left( \hat{a}^{\tR}_{\omega\vec{k}} - e^{-\pi\frac{\omega}{a}}\,\hat{a}^{\tL\dagger}_{\omega\,(-\vec{k})}\right)\,,
\end{equation}
together with $b^\dagger_{\pm \omega,\vec{k}}$. These operators satisfy standard commutation relations 
\begin{equation}
  \left[\hat{b}_{\pm\omega\vec{k}}\,,\hat{b}^\dagger_{\pm\omega\vec{k}} \right] = \delta(\omega-\omega^\prime)\delta(\vec{k}-\vec{k^\prime})\,.
\end{equation}
They are convenient to determine the thermal nature of the Rindler modes in $|0_\tM\rangle$ \cite{Unruh:1976db}. Indeed, using
\begin{equation}
  \aR = \frac{\hat{b}_{-\omega\vec{k}} + e^{-\pi\frac{\omega}{a}}\,\hat{b}^\dagger_{+\omega -\vec{k}}}{\sqrt{1-e^{-2\pi\frac{\omega}{a}}}}\,, \quad
  \aL = \frac{\hat{b}_{+\omega\vec{k}} + e^{-\pi\frac{\omega}{a}}\,\hat{b}^\dagger_{-\omega -\vec{k}}}{\sqrt{1-e^{-2\pi\frac{\omega}{a}}}}\,,
\label{eq:unruh}
\end{equation}
and the fact that Unruh modes annihilate $|0_\tM\rangle$ (eq~\eqref{trind}), one easily finds
\begin{equation}\label{corr}
\begin{aligned}
  \langle 0_\tM\vert\,\aR\aR\,|0_\tM\rangle &=  \langle 0_\tM\vert\,\crR\crR\,|0_\tM\rangle = 0\,, \\
  \langle 0_\tM\vert\,\aR\crR\,|0_\tM\rangle&= \frac{e^{\beta\omega}}{e^{\beta\omega}-1}\,\delta(\omega-\omega^\prime)\delta^{(d-1)}(\vec{k}-\vec{k}^\prime)\,, \\
   \langle 0_\tM\vert\,\crR\aR\,|0_\tM\rangle &= \frac{1}{e^{\beta\omega}-1}\,\delta(\omega-\omega^\prime)\delta^{(d-1)}(\vec{k}-\vec{k}^\prime)\,.
\end{aligned}
\end{equation}
The associated Unruh number and energy operators, labelled with a $U$ superscript, are
\begin{equation}
\begin{aligned}
  \hat{H}^{\tU}&=\int^\infty_0 d\omega\,\int\limits d^{d-1}\vec{k}\,\omega\left[\hat{b}^{\dagger}_{\omega\vec{k}} \hat{b}_{\omega\vec{k}} +  \hat{b}^{\dagger}_{-\omega\vec{k}} \hat{b}_{-\omega\vec{k}}\right]\,, \\
  \hat{N}^{\tU}&= \int^\infty_0 d\omega \int\limits d^{d-1}\vec{k}\left[\hat{b}^{\dagger}_{\omega\vec{k}} \hat{b}_{\omega\vec{k}} +  \hat{b}^{\dagger}_{-\omega\vec{k}} \hat{b}_{-\omega\vec{k}}\right]\,.
\end{aligned}
\end{equation}
It can be verified by direct computation that $N^{\tU}=N^{\tM}$.

Finally, starting from the expression for the boost generator (the total Rindler Hamiltonian in appropriate units)
\begin{equation}\label{Kren}
  \hat{K}^{\textrm{M}}_\text{z} = \frac{1}{a}\left(\hat{H}_\text{R} - \hat{H}_\text{L}\right)\,,
\end{equation}
and using the definition for the Minkowski creation/annihilation operators \eqref{eq:mink-op}, together with the Rindler commutation relations \eqref{eq:right-rindler}, one can verify the
expected commutation relations characteristic of the Poincar\'e algebra
\begin{equation}
  \left[ \hat{H}^{\tM},\hat{K}^{\tM}_\text{z}\right]  = i\hat{P}^{\tM}_{\text{z}}\,, \quad \quad
  \left[\hat{P}^{\tM}_{\text{z}},\hat{K}^{\tM}_\text{z}\right] =- i\hat{H}^{\tM}\,.
\end{equation}

To sum up, using the free algebra of creation and annihilation operators in the left and right Rindler wedges together with the definition of the Minkowski vacuum \eqref{trind}, we constructed a set of operators $\hat{H}^{\tM},\,\hat{P}_{z}^{\tM}$ and $\hat{K}^{\textrm{M}}_\text{z}$, implicitly defined using \eqref{eq:mink-op} and its hermitian conjugate, closing the exact Poincar\'e algebra.



\subsection{From boundary CFT to bulk Poincar\'e algebra}
\label{sec:GFFtoPoincare}

Let us consider large-N holographic theories. Due to large-N factorization, thermal correlation functions are determined by \eqref{eq:exp} and \eqref{eq:N2pa}. As noticed in  \cite{Papadodimas:2012aq}, a normalised version of the Fourier operators
appearing in section~\ref{sec:structure}
\begin{equation}
  \hat{\mathcal{O}}_{\omega,\vec{k}} \equiv \frac{\mathcal{O}_{\omega,\vec{k}}}{\left(G_\beta(\omega,\vec{k})-G_\beta(-\omega,-\vec{k})\right)^{1/2}}= \frac{1}{\sqrt{G_\beta(t,\vec{k})}}\frac{\mathcal{O}_{\omega,\vec{k}}}{\sqrt{1-e^{-\beta\omega}}}\,,
\label{eq:GFFnormal}
\end{equation}
have canonical commutations relations and their thermal expectation values satisfy
\begin{equation}\label{thermal}
\begin{aligned}
  Z_\beta^{-1}\text{Tr}\left(e^{-\beta H}\hat{\mathcal{O}}^\dagger_{\omega,\vec{k}}\hat{\mathcal{O}}_{\omega^\prime,\vec{k}^\prime}\right) &= \frac{1}{e^{\beta\omega}-1}\delta(\omega-\omega^\prime)\delta^{d-1}(\vec{k}-\vec{k}^\prime)\,, \\
  Z_\beta^{-1}\text{Tr}\left(e^{-\beta H}\hat{\mathcal{O}}_{\omega,\vec{k}}\hat{\mathcal{O}}^\dagger_{\omega^\prime,\vec{k}^\prime}\right) &= \frac{e^{\beta\omega}}{e^{\beta\omega}-1}\delta(\omega-\omega^\prime)\delta^{d-1}(\vec{k}-\vec{k}^\prime)\,.
\end{aligned}
\end{equation}
Hence, these operators display the same algebra and expectation values as the right Rindler wedge creation/annihilation operators in \eqref{eq:right-rindler} and \eqref{corr}. We use this observation to explicitly construct boundary CFT operators closing the bulk Poincar\'e algebra, up to $1/N$ corrections, for both 2-sided and 1-sided holographic AdS black holes.

\paragraph{2-sided AdS black holes.}  If the state in the (right) CFT is exactly thermal, we can canonically purify it by the associated thermofield double state
\begin{equation}
  |\text{TFD}\rangle \equiv \frac{1}{\sqrt{Z(\beta)}}\sum_i e^{-\frac{\beta E_i}{2}} |E_i\rangle_{\text{L}}\otimes |E_i\rangle_{\text{R}}
\end{equation}
belonging to the duplicated Hilbert space $\mathcal{H}_{\text{L}} \otimes \mathcal{H}_{\text{R}}$. This purification is precisely the GNS construction of the thermal state reviewed in appendix~\ref{sec:GNS-thermal}. It is holographically dual to the 2-sided eternal AdS black hole \cite{Maldacena:2001kr}. Since the algebra of operators is also duplicated, it gives rise  to two sets of commuting modes $\mathcal{O}^\dagger_{\text{L}\,\omega,\vec{k}}$ and $\mathcal{O}^\dagger_{\text{R}\,\omega,\vec{k}}$ satisfying the same algebra and with the same expectation values as in \eqref{eq:exp} and \eqref{eq:N2pa}. They also verify the following relation 
\begin{equation}\label{tomitO}
 \left(\hat{\mathcal{O}}_{\text{L}\,\omega,\vec{k}} - e^{-\frac{\beta\omega}{2}}\,\hat{\mathcal{O}}^\dagger_{\text{R}\,\omega,\vec{k}}\right)|\text{TFD}\rangle =  \left(\hat{\mathcal{O}}_{\text{R}\,\omega,\vec{k}} - e^{-\frac{\beta\omega}{2}}\,\hat{\mathcal{O}}^\dagger_{\text{L}\,\omega,\vec{k}}\right)|\text{TFD}\rangle = 0\,.
\end{equation}
The origin of this equation is explained in appendix~\ref{sec:GNS-thermal}. It is an special case of eq.~(\ref{tomit2}), and it is basically equivalent to \eqref{trind}, the equation defining $|0_{\text{M}}\rangle$ in the Rindler discussion. As stressed there, our conventions involve time running in the same direction in both wedges. Furthermore, the Rindler acceleration $a$ is mapped to the black hole temperature using $\beta = \frac{2\pi}{a}$.

Given this algebraic equivalence, we can now proceed analogously to our discussion of the different bases of operators and Poincar\'e generators as in Rindler physics. In particular, the generator of time translations in the right CFT reduces in the large-N limit to
\begin{equation}
  \hat{H}^{\text{R}}=\int\limits d\omega\,d\vec{k}\,\omega\,\hat{\mathcal{O}}^\dagger_{\text{R}\,\omega,\vec{k}}\hat{\mathcal{O}}_{\text{R}\,\omega,\vec{k}}\,.
\end{equation}
The Unruh creation/annihilation operators can be defined by
\begin{equation}\label{eq:unruh}
\mathcal{O}^{\textrm{U}}_{+\omega,\vec{k}}= \frac{1}{\sqrt{1-e^{-\beta\omega}}}\left( \hat{\mathcal{O}}_{\text{L}\,\omega,\vec{k}} - e^{-\beta\omega/2}\,\hat{\mathcal{O}}^\dagger_{\text{R}\,\omega,\vec{k}}\right) \,, \quad
\mathcal{O}^{\textrm{U}}_{-\omega,\vec{k}}=\frac{1}{\sqrt{1-e^{-\beta\omega}}}\left( \hat{\mathcal{O}}_{\text{R}\,\omega,\vec{k}} - e^{-\beta\omega/2}\,\hat{\mathcal{O}}^\dagger_{\text{L}\,\omega,\vec{k}}\right) \;,
\end{equation}
while Minkowski annihilation modes can be defined by
\begin{equation}
  \mathcal{O}^\tM_{k_z\vec{k}} = \int^\infty_0 \frac{d\omega}{\sqrt{2\pi ak_0}}\frac{1}{\sqrt{1-e^{-\beta\omega}}} \left[e^{i\,2\pi\vartheta(k_z)\omega/\beta} \left(\hat{\mathcal{O}}_{\text{L}\,\omega,\vec{k}} - e^{-\beta\omega/2}\,\hat{\mathcal{O}}^\dagger_{\text{R}\,\omega,\vec{k}}\right) + e^{-i\,2\pi\vartheta(k_z)\omega/\beta} \left(\hat{\mathcal{O}}_{\text{R}\,\omega,\vec{k}} - e^{-\beta\omega/2}\,\hat{\mathcal{O}}^\dagger_{\text{L}\,\omega,\vec{k}}\right)\right]\,.
\end{equation}
These operators allow us to define operators generating Minkowski time and spatial $z$ translations by
\begin{eqnarray}
\hat{H}^{\tM}&=&\int\limits dk_{z}\,  d^{d-1}\vec{k}\,\omega_{\vec{k}}\,\mathcal{O}^{\tM\dagger}_{k_z\vec{k}}\mathcal{O}^\tM_{k_z\vec{k}}\nonumber\\
\hat{P}_{z}^{\tM}&=&\int\limits  dk_{z}\,d^{d-1}\vec{k}\,k_{z}\,\mathcal{O}^{\tM\dagger}_{k_z\vec{k}}\mathcal{O}^\tM_{k_z\vec{k}}\;,
\end{eqnarray}
where $\omega_{\vec{k}}^{2}=m^{2}+k_{z}^{2}+\vert\vec{k}\vert^{2}$. The associated number operator is just given by
\begin{equation}
\hat{N}^{\textrm{M}}=\int\limits  dk_{z}\,d^{d-1}\vec{k}\,\mathcal{O}^{M\dagger}_{k_z\vec{k}}\mathcal{O}^M_{k_z\vec{k}}\;.
\end{equation}
Since the algebra of the modes $\hat{\mathcal{O}}_{\text{R}\,\omega,\vec{k}}$ and $\hat{\mathcal{O}}_{\text{L}\,\omega,\vec{k}}$, together with the expectation values in the thermofield double, are equal to the ones in the Rindler discussion, up to $1/N$ corrections, we conclude these boundary CFT operators close the same bulk Poincar\'e algebra
\begin{equation}\label{pal}
\left[ \hat{H}^{\tM},\hat{H}^{\textrm{T}}\right]  = i\hat{P}^{\tM}_{z}\,, \quad
\left[\hat{P}^{\tM},\hat{H}^{\textrm{T}}\right]  =- i\hat{H}^{\tM}\,,
\end{equation}
where the boost operator $\hat{H}^{\textrm{T}}$ is the total boost Hamiltonian of the two decoupled CFTs defined by
\begin{equation}\label{Hren}
\hat{H}^{\textrm{T}}=\frac{\beta}{2\pi}\left(\hat{H}^\textrm{R} - \hat{H}^\textrm{L}\right)
\end{equation}
This is the same as in the Rindler discussion~\eqref{Kren}, after the replacement $a\to 2\pi/\beta$.

\paragraph{1-sided AdS black holes.} The work in \cite{Papadodimas:2012aq,Papadodimas:2013wnh,Papadodimas:2013jku} extends the previous discussion to single AdS black holes involving a single boundary CFT.  Observing that energy eigenstates $|E_i\rangle$ are well approximated by the thermal ensemble, one can work within the code subspace \cite{Almheiri:2014lwa,Harlow:2018fse}
and show that $|E_i\rangle$ is a \emph{cyclic} and a \emph{separating} vector in the Hilbert space with respect to the code subspace algebra \cite{deBoer:2018ibj,deBoer:2019kyr}. The Tomita-Takesaki theorem\footnote{See the book \cite{Haag:1992hx} for a physics introduction, the summary done in Ref. \cite{Papadodimas:2013wnh}, or Ref. \cite{Witten:2018lha} for a recent review.} guarantees the existence of a non-trivial ``mirror'' commutant. The mirror operators $\tilde{O}_{\omega,\vec{k}}$ generating this commutant play a similar role to the operators $\mathcal{O}_{\text{L}\,\omega,\vec{k}}$ in the 2-sided discussion, but they belong to the same boundary CFT dual to the single AdS black hole. They are defined by the following relations
\begin{equation}
\begin{aligned}
  \tilde{\mathcal{O}}_{\omega,\vec{k}} |\Psi_i\rangle &= e^{-\frac{\beta\omega}{2}}\,\mathcal{O}^\dagger_{\omega,\vec{k}}\,  |\Psi_i\rangle \,, \\
  \tilde{\mathcal{O}}_{\omega,\vec{k}} \mathcal{O}_{\omega_1,\vec{k}_1}\dots \mathcal{O}_{\omega_n,\vec{k}_n} |\Psi_i\rangle  &= \mathcal{O}_{\omega_1,\vec{k}_1}\dots \mathcal{O}_{\omega_n,\vec{k}_n} \tilde{\mathcal{O}}_{\omega,\vec{k}} |\Psi_i\rangle \,, \\
  \left[H, \tilde{\mathcal{O}}_{\omega,\vec{k}}\right] \mathcal{O}_{\omega_1,\vec{k}_1}\dots \mathcal{O}_{\omega_n,\vec{k}_n} |\Psi_i\rangle  &= \omega\,\tilde{\mathcal{O}}_{\omega,\vec{k}}  \mathcal{O}_{\omega_1,\vec{k}_1}\dots \mathcal{O}_{\omega_n,\vec{k}_n} |\Psi_i\rangle\,.
\end{aligned}
\label{eq:mirror-def}
\end{equation}
These relations should be understood to hold only within the code subspace. The mirror operators are thus state dependent. As discussed in \cite{deBoer:2019kyr}, there are some ambiguities regarding the $1/N$ extension of these operators, but for us it will be enough to work within the code subspace. This, together with microstates $|E_i\rangle$ being well approximated by the canonical ensemble, ensures the defining properties \eqref{eq:mirror-def} give rise to an algebra and correlation functions that are equivalent to the algebra of the $\mathcal{O}_{\text{L}\,\omega,\vec{k}}$ in the 2-sided discussion, and therefore to the algebra of annihilation operators in the left wedge. The previous formulas showing the existence of a Poincar\'e algebra can just be extended to this situation by the replacement $\mathcal{O}_{\text{L}\,\omega,\vec{k}}\rightarrow \tilde{O}_{\omega,\vec{k}}$.

\section{Growth measures}
\label{sec:growth}

As argued in section~\ref{sec:structure}, time evolution of simple perturbations in large-N theories at any finite temperature is simply described in terms of Fourier modes $\mathcal{O}_{\omega,\vec{k}}$ and $\mathcal{O}^\dagger_{\omega,\vec{k}}$. These modes allow for the evaluation of the series of nested commutators \eqref{lno} in any energy eigenstate compatible with the given temperature if the theory satisfies the ETH conjecture. Within the large-N limit,\footnote{Our discussion focuses on large-N theories, but it also applies to free QFT in Rindler space given the algebraic equivalence between Rindler operators and the Fourier modes $\mathcal{O}_{\omega,\vec{k}}$ and $\mathcal{O}^\dagger_{\omega,\vec{k}}$, as explicitly discussed in section~\ref{sec:algebra}. To our knowledge, operator growth in Rindler space has not been considered in the literature and it is useful to gauge away some of the confusions arising when defining the notion of operator growth in QFT.} this operator structure should be enough to characterize any notion of operator growth.

In this section, we show this last expectation is indeed the case by describing some existent notions of operator growth, such as operator size in spin systems or the recursion method in many-body physics. We finish with some discussion regarding the relation between Nielsen's geometric formulation of quantum circuit complexity, operator growth and quantum/classical chaos. In the way, we provide a generic framework to study operator growth in QFT, based on the GNS construction reviewed in appendix~\ref{App2}. Albeit our applications will be confined to large-N theories, the present GNS approach might help to understand the putative definitions of operator growth in generic QFT's.

\subsection{Operator growth as state mixing in the GNS construction}
\label{sec:GNS}

Talking about operator growth requires to be able to expand a given operator in different bases of the space of operators, to quantify how the support of the operator changes with time. Hence, given an operator algebra $\mathcal{A}$, we need an inner product endowing $\mathcal{A}$ with the structure of a Hilbert space. This is precisely the goal of the GNS construction, which is reviewed in appendix~\ref{App2}. Here we briefly summarize its main ingredients. Given a state\footnote{The word \emph{state} refers to a linear functional acting on the algebra $\mathcal{A}$ as properly defined in \eqref{eq:omega-state}. This is the standard terminology used in algebraic QFT \cite{Haag:1992hx}.} $\phi$ acting on the algebra $\mathcal{A}$ satisfying
\begin{equation}
A\in \mathcal{A}\, ,\,\,\,\,\, \phi (A^{\dagger}A) =0 \quad \Longleftrightarrow \quad A=0\,,
\end{equation}
the GNS Hilbert space $\mathcal{H}_{\phi}$ and its inner product are defined by
\begin{equation}
  A\in\mathcal{A}\Rightarrow \vert A\rangle \in \mathcal{H}_{\phi}\,, \quad \langle B\vert A\rangle \equiv \phi (B^{\dagger}A) \,.
\end{equation}
In this Hilbert space there are two equivalent representations $\pi$ and $\bar{\pi}$ of the algebra $\mathcal{A}$ acting on $\mathcal{H}_{\phi}$
\begin{equation}
A\in\mathcal{A}\Rightarrow \vert A\rangle \in \mathcal{H}_{\phi}\,, \quad \pi (A)\vert B\rangle \equiv \vert AB\rangle \, , \,\,\,\,\,\, \bar{\pi} (A)\vert B\rangle \equiv \vert BA^{\dagger}\rangle\,.
\end{equation}
This construction is valid for any type of operator algebras, including the type III algebras relevant for QFT.

Consider states $|\kappa\rangle$ arising from abstract states in the algebra which are invariant under time evolution, such as thermofield double states.\footnote{Notice that this can be straightforwardly generalized to states invariant under so-called modular time evolution. Also, notice that this definition is not restricted to time-independent states. Starting with the invariant one we can move to other states by using elements of the algebra. These states would then evolve as it is described.} In this context, the GNS construction maps the Heisenberg time evolution $A(t)$ of any operator belonging to the algebra to the Schr\"odinger's time evolution of the associated GNS state
\begin{equation}
  U(t) \pi(A)\vert \kappa\rangle\equiv \pi (A(t))\vert \kappa\rangle =\vert A(t)\kappa\rangle \equiv \vert \Psi (t)\rangle\,,
\end{equation}
The same conclusion holds for the representation $\bar{\pi}$.

Since operator evolution is equivalent to state evolution in the associated GNS Hilbert space, any notion of operator growth should be characterized by expectation values of \emph{size} operators in the GNS Hilbert space $\mathcal{H}_{\phi}$
\begin{equation}\label{gen}
\langle \Psi (t)\vert \sum\limits_{ij}\bar{\pi}(B_i) \pi (B_j)\vert \Psi (t)\rangle\,,
\end{equation}
as any other property attached to states in $\mathcal{H}_{\phi}$. In the previous relation $B_j$ runs over a basis of operators of the algebra. Below we show how this is the case in some recent examples in lattice systems.

\paragraph{Operator size for simple Majorana operators.} Before exploring this perspective for large-N theories, we briefly comment on how the case of Majorana spin systems at infinite temperature  \cite{Roberts:2018mnp} and its extension to finite temperature  \cite{Qi:2018bje} fit in this framework. 

Consider a set of $N$ fundamental Majorana operators normalized by $\lbrace\psi_{a},\psi_{b}\rbrace^{2}=2\delta_{ab}$. Every operator $\psi$ in the algebra $\mathcal{A}$ can be expanded as
\begin{equation}
\mathcal{O} =\sum\limits_{s=1}^{N}\,\sum\limits_{a_{1}\cdots a_{s}}c_{a_{1}\cdots a_{s}}\psi_{a_{1}}\cdots \psi_{a_{s}}\;.
\end{equation}
The \emph{size} of such operator was defined by \cite{Roberts:2018mnp} 
\begin{equation}\label{size}
S_{\mathcal{O}}=\sum\limits_{s=1}^{N}\,s\sum\limits_{a_{1}\cdots a_{s}}\vert c_{a_{1}\cdots a_{s}}\vert^2\,.
\end{equation}
This is natural if one thinks of the label $s$ as describing, either location in a 1d lattice, or directly in terms of the number of fundamental fermions building the operator. 

The connection to the GNS construction is as follows. One assigns a vector $|\mathcal{O} \rangle \in \mathcal{H}_{\phi}$ to every operator $\mathcal{O} \in \mathcal{A}$ with GNS inner product
\begin{equation}
\langle\mathcal{O} \vert\mathcal{O}'\rangle\equiv \frac{1}{Z}\textrm{Tr}(\mathcal{O}^{\dagger}\mathcal{O}')\;,
\end{equation}
where $Z$ is the dimension of the Hilbert space $Z=\textrm{Tr}(\mathds{1})$ induced by the normalized inner product for finite matrices. It follows the identity element of the algebra $\mathds{1}\to \vert\mathds{1}\rangle$ and the inner product can be interpreted as an expectation value at infinite temperature. There is one natural representation of the algebra in $ \mathcal{H}_{\phi}$, defined by $\pi (\mathcal{O})\vert\mathcal{O}'\rangle =\vert\mathcal{O}\mathcal{O}'\rangle$. This allows to define unitary time evolution in the GNS Hilbert space by
\begin{equation}
U_{t}\pi(\mathcal{O})\vert\mathds{1}\rangle\equiv\pi(\mathcal{O}(t))\vert\mathds{1}\rangle =\vert\mathcal{O}(t)\rangle \;.
\end{equation}
This gives a concrete example on how operator evolution is seen as state evolution in the GNS Hilbert space.

Within the GNS construction, there exists an operator $\hat{S}$ acting on $\mathcal{H}_{\phi}$
\begin{equation}\label{s1}
\hat{S}=\sum\limits_{s=1}^{N}\,s \sum\limits_{a_{1}\cdots a_{s}}\pi (\psi_{a_{1}}\cdots \psi_{a_{s}})\vert\mathds{1}\rangle\langle\mathds{1}\vert \pi (\psi_{a_{1}}\cdots \psi_{a_{s}})= \sum\limits_{s=1}^{N}\,s \sum\limits_{a_{1}\cdots a_{s}}\vert\psi_{a_{1}}\cdots \psi_{a_{s}}\rangle\langle\psi_{a_{1}}\cdots \psi_{a_{s}}\vert\;,
\end{equation}
satisfying
\begin{equation}
S_{\mathcal{O}}(t)=\langle\mathcal{O} (t)\vert \hat{S}\vert \mathcal{O} (t)\rangle\,.
\end{equation}
Hence, the notion of size \eqref{size} equals the expectation value of $\hat{S}$ in the GNS state $\vert \mathcal{O} (t)\rangle$ that is mapped to the time evolution of the original operator $\mathcal{O}(t)$. This matches our general expectation that any notion of operator growth should be computable by expectation values evaluated on the GNS state. In fact, introducing creation $c^{\dagger}_{i}$ and annihilation $c_{i}$ operators in $\mathcal{H}_{\phi}$ by 
\begin{equation}
\begin{aligned}
  c_{i}\vert\psi_{a_{1}}\cdots \psi_{a_{s}}\rangle &=\delta_{i,a_{1}}\vert\psi_{a_{2}}\cdots \psi_{a_{s}}\rangle +\cdots +\delta_{i,a_{s}}\vert\psi_{a_{2}}\cdots \psi_{a_{s-1}}\rangle\,, \\
  c^{\dagger}_{i}c_{i}\vert\psi_{a_{1}}\cdots \psi_{a_{s}}\rangle &=\delta_{i,a_{1}}\cdots \delta_{i,a_{s}}\,,
\end{aligned}
\label{s2}
\end{equation}
the size operator \eqref{s1} can be reinterpreted as a \emph{number} operator
\begin{equation}
  \hat{S}=\sum\limits_{i=1}^{N}c^{\dagger}_{i}c_{i}\;.
\label{eq:f-size}
\end{equation}
Hence, we learn there is a basis of operators in $\mathcal{H}_{\phi}$ where a natural notion of size in this lattice system is a simple quadratic operator.

This notion of size in lattice systems was extended to finite temperature in \cite{Qi:2018bje} by purifying the thermal ensemble of the Majorana fermions. This approach is then directly on a GNS form, as reviewed in appendix~\ref{sec:GNS-thermal}, except that using fermionic operators. Hence, our conclusions extend to this finite temperature case too.

\paragraph{Alternative notions of operator size} Depending on the dynamics and state of the system, operator size defined as in \eqref{size} may not be a dynamical quantity. This can happen even if the complexity of the operator grows. Indeed, if the Hamiltonian preserves the number of particles, the previous definition of operator size will be a conserved charge for a natural class of initial states, as we review now.

Consider a bunch of spinless fermions whose hamiltonian conserves the number of particles. Any state can be expanded as
\begin{equation}
\vert \psi\rangle =\sum\limits_{s=0}^{N}\,\sum\limits_{a_{1}\cdots a_{s}}\psi_{a_{1}\cdots a_{s}}c^{\dagger}_{a_{1}}\cdots c^{\dagger}_{a_{s}}\vert \downarrow_{1}\cdots \downarrow_{N}\rangle \;,
\end{equation}
where $c^{\dagger}_{i}$ and $c_{i}$ create and destroy such fermions at site $i$. Besides operator size, one can ask about how many particles are being transported by the Hamiltonian as time evolves. This was considered in \cite{Magan:2016ehs}. To be definite, consider an starting state with the first $m$ particles excited. Unitary evolution mixes the state with other states in the $m$-particle sector
\begin{equation}
\vert \psi (t)\rangle =\sum\limits_{a_{1}\cdots a_{m}}\psi_{a_{1}\cdots a_{m}} (t)\,c^{\dagger}_{a_{1}}\cdots c^{\dagger}_{a_{m}}\vert \downarrow_{1}\cdots \downarrow_{N}\rangle\;.
\end{equation}
One proposed notion of operator growth is the average number of jumps (the average transport) the spins have performed due to Hamiltonian evolution. Such number can be measured by the expectation value of the number operator
\begin{equation}\label{n1}
\hat{T}\equiv\sum\limits_{i=m+1}^{N}c^{\dagger}_{i}c_{i}
\end{equation}
counting the number of fermions in the sites that were not populated at $t=0$. It follows
\begin{equation}
T(t)\equiv\langle\psi (t)\vert  \hat{T}\vert\psi (t)\rangle
\end{equation}
is the average particle transport. This is a sensible measure of the growth of the state $\vert\psi (t)\rangle$ and it is still the expectation value of a simple operator. Also, since
\begin{equation}
\vert \psi (t)\rangle = U(t)c^{\dagger}_{1}\cdots c^{\dagger}_{m}U^{-1}(t)\vert \downarrow_{1}\cdots \downarrow_{N}\rangle \equiv \mathcal{O}(t) \vert \downarrow_{1}\cdots \downarrow_{N}\rangle\;,
\end{equation}
this expectation value is equivalently studying the growth of the operator $\mathcal{O}=c^{\dagger}_{1}\cdots c^{\dagger}_{m}$ in the state $\vert \downarrow_{1}\cdots \downarrow_{N}\rangle $. Notice that while this notion of size is bound to grow, as analized in \cite{Magan:2016ehs}, the previous notion of size, where we would add up all spins in the definition~(\ref{n1}), would be constant through time evolution due to particle number conservation.

\subsection{Size, number operators and energies in large-N and holographic theories}
\label{sec:size-N}

The observation that Heisenberg operator evolution is equivalent to Schr\"odinger's time evolution in the GNS Hilbert space provides a hint to extend the notion of operator size to QFT. 
Such notion is based on \emph{simple} operators, such as \eqref{eq:f-size} and \eqref{n1} in spin systems (see \cite{Magan:2016ehs,Roberts:2018mnp,Qi:2018bje,Parker:2018yvk}). 

Given the structure of time evolution for holographic CFTs in the large-N limit discussed in sections~\ref{sec:structure} and \ref{sec:GFFtoPoincare}, 
it may be natural to define any notion of operator size as being of the form
\begin{equation}\label{sizeQFT}
  \hat{S}=\sum_\alpha F_{\alpha}\left[\left(\mathcal{O}_\alpha\right)_{\omega,\vec{k}},\left(\mathcal{O}^\dagger_\alpha\right)_{\omega,\vec{k}},
  \left(\tilde{\mathcal{O}}_\alpha\right)_{\omega,\vec{k}},\left(\tilde{\mathcal{O}}^\dagger_\alpha\right)_{\omega,\vec{k}}\right]
\end{equation}
in terms of the operator modes and its mirror partners for the different local low conformal dimension boundary operators indexed by $\alpha$. The additive nature on the spectrum of operators is due to the absence of mixing between operators when neglecting 1/N corrections.


Assuming the generic definition \eqref{sizeQFT}, large-N factorization ensures that any notion of size in large-N theories associated with a choice of the functionals $F_\alpha$ is completely determined by the two-point function~(\ref{eq:exp}), up to 1/N corrections. Indeed, to study the growth of the size of a certain field $\mathcal{O}(t)$ in the thermofield double $\vert\kappa_{\beta}\rangle$ at temperature $\beta$, we just need to take the expectation value of $\hat{S}$ in the evolving GNS state $\vert\mathcal{O}(t)\kappa_{\beta}\rangle$. Such expectation value can be computed by using expressions~(\ref{exparep}) and~(\ref{eq:exp}), together with large-N factorization.

Let us remark that in previous literature, starting with \cite{Roberts:2018mnp}, the notion of size has been argued to be related to out-of-time ordered correlation functions. If this relation is extended to any large-N theory and any temperature, our analysis would imply a non-trivial relation between two and four-point functions in the large-N limit. These aspects will be studied elsewhere.

In analogy to the spin size \eqref{eq:f-size}, natural choices for the size $\hat{S}$ in large-N QFTs are simple operators like the number or energy operators associated to the different bases (Rindler, Unruh, and Minkowski) of creation and annihilation operators discussed in section~\ref{sec:GFFtoPoincare}. As in our discussion of the dynamics preserving the number of particles, not all choices for such operators will provide useful dynamical information. In what follows, the upper index in the size operator refers to the basis chosen, either Rindler (R), Unruh (U) or Minkowski (M), and the lower index to whether it is energy-based (H) or particle number based (N).  

Let us start our discussion with the number and energy operator associated to the standard basis of operator modes
\begin{equation}
   \hat{S}^{\textrm{R}}_{\textrm{N}} = \hat{N}=\int\limits_{\omega> 0}d\omega  d^{d-1}\vec{k}\, \mathcal{O}^{\dagger}_{\omega,\vec{k}}\mathcal{O}_{\omega,\vec{k}}\, \quad \text{and} \quad
   \hat{S}^{\textrm{R}}_{\textrm{H}} = \hat{H}=\int\limits_{\omega> 0}d\omega  d^{d-1}\vec{k}\,\,\omega\,\mathcal{O}^{\dagger}_{\omega,\vec{k}}\mathcal{O}_{\omega,\vec{k}}\,.
\end{equation}
Then the expectation value $\langle \kappa_\beta\mathcal{O}(t)|\hat{S}|\mathcal{O}(t)\kappa_\beta\rangle$ is constant for any operator $\mathcal{O}$ and provides no further dynamical information.

\paragraph{Unruh and Minkowski number operators.}  Consider the Unruh and Minkowski number operators choice
\begin{equation}
  \hat{S}^{\textrm{U}}_{\textrm{N}}= \hat{N}^{\textrm{U}}= \int\limits dk_{z}\,  d^{d-1}\vec{k}\,\mathcal{O}^{\textrm{U}\dagger}_{\omega\vec{k}}\mathcal{O}^\textrm{U}_{\omega\vec{k}}\,, \quad \text{and} \quad
  \hat{S}^{\textrm{M}}_{\textrm{N}}=  \hat{N}^{\textrm{M}}= \int\limits dk_{z}\,  d^{d-1}\vec{k}\,\mathcal{O}^{\textrm{M}\dagger}_{k_z\vec{k}}\mathcal{O}^\textrm{M}_{k_z\vec{k}}\,.
\end{equation}
Both choices are equal due to the algebra of field modes. This proposal is inspired by the relation~(\ref{tomitO}), which is a specific instance of the more general Tomita-Takesaki like equation \eqref{tomit2}. One basically defines the simplest operators annihilating the thermofield double\footnote{Notice there is an infinite number of choices that also annihilate the thermofield double, including the Minkowski annihilation/creation operators.} and uses them to define a number operator. For Majorana fermions this was the path chosen in \cite{Qi:2018bje}. Here we see that such operators, in the black hole scenario, are the known Unruh creation/annihilation operators.

These notions of size give zero on the thermofield double, while positive sizes on the thermofield double with perturbations. However, if we consider a Minkowski mode excitation generated by $\mathcal{O}^{\textrm{M}\dagger}_{k_z\vec{k}}$, this notion of size will remain constant through time evolution. This is because unitary evolution acts like a boost on these excitations. Hence, it changes the value of their momentum but not the number of modes.  This choice shows that depending on the basis of operators being considered, equating size with number operators may not provide useful dynamical information.

We remark that although the Minkowski creation/annihilation operators are state-dependent in the one-sided case since they make use of the mirror operators, exciting a Minkowski mode is not a state-dependent action. The reason is that the action of the mirror creation operator can be fully mimicked by their partners in the right wedge, as in \eqref{ami3}.

\paragraph{Minkowski energy.} Consider the boundary CFT operator describing the analog of a bulk infalling Hamiltonian\footnote{Here we are referring to the Minkowski energy choice. The Unruh energy is not a sensible choice since it is infinite for any Minkowski mode.} constructed in the previous section
\begin{equation}
  \hat{S}^{\textrm{M}}_{\textrm{H}}= H^{\textrm{M}}= \int\limits dk_{z}\,  d^{d-1}\vec{k}\,\omega_{\vec{k}}\,\mathcal{O}^{\textrm{M}\dagger}_{k_z\vec{k}}\mathcal{O}^{\textrm{M}\dagger}_{k_z\vec{k}}\,.
\end{equation}
This choice was studied in detail in \cite{Magan:2018nmu}. Since the CFT time evolution acts as a boost upon a Minkowski mode $\mathcal{O}^{\textrm{M}\dagger}_{k_z\vec{k}}$
\begin{equation}
  e^{i\gamma\,K}\,\mathcal{O}^{\textrm{M}\dagger}_{k_z\vec{k}}\,e^{-i\gamma\,K} = \frac{\sqrt{\gamma(\omega_{\vec{k}})}}{\sqrt{\omega_{\vec{k}}}}\,\mathcal{O}^{\textrm{M}\dagger}_{\gamma(k_z)\vec{k}}\,,
\end{equation}
it follows the size will exponentially grow at large times $(t^\text{M}\gg \beta)$, with Lyapunov exponent equal to $2\pi/\beta$, since evolving with the QFT Hamiltonian for time $t$ is equivalent to boosting the particle with rapidity $\frac{2\pi}{\beta} t$ (see~\eqref{Hren}).


\subsection{The recursion method at large-N}
\label{sec:recursion}

A standard approach in condensed matter physics to study the Heisenberg time evolution and complexity of operators is the recursion method (see \cite{viswanath2008recursion} for a detailed presentation). This perspective on operator growth was recently considered in SYK in \cite{Parker:2018yvk} and used to study long time scales of operator dynamics in \cite{Barbon:2019wsy}. In this section  we explore what the structure of time evolution in large-N theories described in sections~\ref{sec:structure} and \ref{sec:GFFtoPoincare} teaches us about this approach.

Let us first describe this method briefly. As before, the recursion method requires the definition of an inner product. The book \cite{viswanath2008recursion} considers a whole family of them, but we show in appendix~\ref{inner-products} that all choices can be related to one convenient representative in QFT. In the following we focus on such representative and, for simplicity, we only consider operators with vanishing one-point functions. Given two operators $A$ and $B$, the representative inner product $(A,B)$ is defined by
\begin{equation}
  (A,B)\equiv \langle e^{\beta H/2}A^{\dagger}e^{-\beta H/2} B\rangle_{\beta}\,,
\label{innerL}
\end{equation}
where $\langle A \rangle_\beta = \text{Tr}(\rho_\beta\,A)$. Within the GNS framework, this inner product can be written as
\begin{equation}
  \langle \kappa\vert \bar{\pi}(A)\pi (B)\vert \kappa\rangle =\langle \kappa\vert B\kappa A^\dagger\rangle=\langle e^{\beta H/2}A^{\dagger}e^{-\beta H/2} B\rangle_{\beta}=(A,B) \,.
\label{eq:inner-GNS}
\end{equation}

The inner product \eqref{innerL} allows to expand any hermitian operator $\mathcal{O}(t)$ in an orthogonal basis of operators
\begin{equation}\label{exp}
  \mathcal{O}(t)=\sum\limits_{k=0}^{\infty}C_{k}(t)f_{k}\,, \quad \text{with} \quad (f_{k},f_{k'})=(f_{k},f_{k})\,\delta_{kk'}
\end{equation}
for some time dependent coefficients $C_{k}(t)$. The recursion method proposes to use an explicit basis $f_{k}$, the Lanczos basis, to study operator evolution. Basically, starting from the operators $\mathcal{O}_{n}$ defined previously, it provides a constructive algorithm, based on the Gram-Schmidt orthogonalization procedure, to determine such basis
\begin{equation}
  f_{k+1} = i[H,\,f_k] + \Delta_k\,f_{k-1}\,, \quad k=0,1,\dots \quad \text{with} \quad \Delta_k = \frac{\left(f_j,\,f_j\right)}{\left(f_{j-1},\,f_{j-1}\right)}\,, \,\,\,j=1,2,\dots
\label{eq:rec-basic}
\end{equation}
with initial conditions $f_{-1}=0$ and $f_0=\mathcal{O}$ being the operator at initial time. The coefficients $\Delta_k$ are referred to as the Lanczos coefficients. The linear operator $\mathcal{L}$ generating the time evolution as $\mathcal{L}(\mathcal{O})\equiv [H,\mathcal{O}]$ is sometimes called the Liouvillian. It corresponds to the GNS hamiltonian in the GNS construction described in appendix \ref{App2} generating the unitary evolution in \eqref{ugns1}-\eqref{ugns2}.

Plugging the expansion \eqref{exp} into Heisenberg's equation of motion, one derives a 1d diffusion equation for the amplitudes $C_k(t)$
\begin{equation}
  \frac{d\mathcal{O}}{dt} = i\mathcal{L}\,\mathcal{O} \quad \Leftrightarrow \quad \dot{C}_k(t) = C_{k-1}(t) - \Delta_{k+1}\,C_{k+1}(t)\,, \quad k=0,1,2\dots
\label{eq:diffusion}
\end{equation}
with initial conditions $C_{-1}(t)=0$ and $C_k(0) = \delta_{k0}$. In this framework, the Lanczos operator complexity $L_{\mathcal{O}}$ of an operator $\mathcal{O}$ can be defined by the average position in this effective 1d chain\footnote{When adopting this definition, one is assigning some kind of locality interpretation to the 1d chain which was manifest in our Majorana fermion discussion \eqref{size}. Here, the label $k$ technically accounts for the number of commutators with $H$ that have acted upon the starting operator. Thus, depending on the interactions in this hamiltonian, the support of the different $f_k$ operators will grow accordingly. Since one is interested in quantifying the evolution in the size of this support, one could consider more general functionals $\sum\limits_{k=0}^{\infty}g(k)\,\frac{\vert (f_{k},\mathcal{O}(t))\vert^{2}}{(f_{k},f_{k})}$ capturing this quantity more precisely. Our point here is that the existent inner product allows to define a notion of average size for any relevant choice of $g(k)$.}
\begin{equation}
  L_{\mathcal{O}}\equiv \sum\limits_{k=0}^{\infty}k\,\frac{\vert (f_{k},\mathcal{O}(t))\vert^{2}}{(f_{k},f_{k})}\,.
\label{lc}
\end{equation}
Because of \eqref{eq:inner-GNS}, $L_{\mathcal{O}}$ equals the expectation value of the ``Lanczos operator'' in the GNS Hilbert space
\begin{equation}
  \hat{L}_{\mathcal{O}}\equiv\sum\limits_{k=0}^{\infty}k\, \frac{\bar{\pi}(f_{k})}{\sqrt{(f_{k},f_{k})}}\vert\kappa\rangle\langle \kappa\vert \frac{\bar{\pi}(f_{k})}{\sqrt{(f_{k},f_{k})}}\,,
\end{equation}
so that
\begin{equation}
L_{\mathcal{O}}=\langle \kappa\vert \pi (\mathcal{O}(t)) \hat{L}_{\mathcal{O}}\pi (\mathcal{O}(t))\vert\kappa\rangle\,.
\end{equation}
Notice $\hat{L}_{\mathcal{O}}$ is a state dependent operator in the GNS Hilbert space, i.e. it is \emph{not} linear in $\mathcal{O}$. 

Knowledge of the Lanczos coefficients $\Delta_n$ determines the orthogonal basis $\{f_k\}$ and allows to determine the full-time evolution of the operator $\mathcal{O}(t)$ through the integration of the 1d diffusion equation \eqref{eq:diffusion}. Interestingly, these coefficients $\Delta_n$ can be recursively extracted from the connected 2-pt function
\begin{equation}
  Q(t) \equiv \langle \mathcal{O}(t)\mathcal{O}(0)\rangle_\beta \,.
\label{eq:2ptbeta}
\end{equation}
This can be seen as follows \cite{viswanath2008recursion}. Since
\begin{equation}
  C_{0}(t) \equiv \frac{\left(\mathcal{O}(t),\,\mathcal{O}\right)}{\left(\mathcal{O},\,\mathcal{O}\right)} = \frac{\Phi(t)}{\Phi(0)}\,,
\end{equation}
where $\Phi(t) \equiv \left(Q(t) + Q(-t)\right)/2$, if one assumes the Taylor expansion
\begin{equation}
  C_0(t) = \sum_{k=0}^\infty \frac{(-1)^k}{(2k)!}\,M_{2k}\,t^{2k}\,,
\end{equation}
is sensible, it follows the existence of a one--to--one reconstruction algorithm between the moments $M_{2k}$, which are determined solely by derivatives of the 2-pt function \eqref{eq:2ptbeta}, and the Lanczos coefficients $\Delta_k$
\begin{equation}
  M_{2k}^{(n)} = \frac{M_{2k}^{(n-1)}}{\Delta_{n-1}} - \frac{M_{2k-2}^{(n-2)}}{\Delta_{n-2}}\,, \quad \Delta_n=M_{2n}^{(n)}\,, \quad k=n,n+1,\dots ,K\,\,\,\,\,\text{and} \,\,\,\, n=1,2,\dots ,K
\label{eq:moment-Delta}
\end{equation}
The initial conditions of this recursion are $M_{2k}^{(0)}=M_{2k}$ and $\Delta_{-1}=\Delta_0=1$ and $M_{2k}^{(-1)}=0$.

In section~\ref{sec:structure}, the structure of time evolution in large-N theories was discussed. It is natural to ask whether we can learn anything about the recursion method given this structure. The first observation is that the operators $\mathcal{O}_{\omega,\vec{k}}$ diagonalize the Liouvillian
\begin{equation}
  \frac{d \mathcal{O}_{\omega,\vec{k}}(t)}{dt}=i\mathcal{L} (\mathcal{O}_{\omega,\vec{k}})=i[H,\mathcal{O}_{\omega,\vec{k}}]=-i\omega \,\mathcal{O}_{\omega,\vec{k}}\quad \Longrightarrow \quad \mathcal{O}_{\omega,\vec{k}}(t)=e^{-i\omega t}\,\mathcal{O}_{\omega,\vec{k}}(t)
\end{equation}
Hence, there are two natural bases in the space of operators: the Lanczos basis, made of the orthogonal $f_k$ and the operator modes $\mathcal{O}_{\omega,\vec{k}}$ diagonalising the Liouvillian in the large-N limit. Finding the change of basis would immediately solve the diffusion equation \eqref{eq:diffusion}, since the time evolution of the $\mathcal{O}_{\omega,\vec{k}}$ operators is known. Since properly normalised operator modes \eqref{eq:GFFnormal} satisfy the canonical commutation relation associated to a set of of creation and annihilation operators, we analyse the latter from the recursion method perspective next.

Consider $K$ free harmonic oscillators, denoted by $a_{j}$ and $a^\dagger_j$, with $j=1,\dots ,K$. Take as our starting operator in the recursion method algorithm
\begin{equation}
  f_0 = \sum_j D_j (a_j+a^\dagger_j)\,, \quad \quad D_i\in \mathbb{R}
\end{equation}
with $D_j$ any set of coefficients. This choice matches the $t=0$ expansion of the field in \eqref{exparep} and accommodates their normalisation \eqref{eq:GFFnormal}. Since $\mathcal{L}a^\dagger_j = \omega_j\,a^\dagger_j$ and $\mathcal{L} a_j = -\omega_j\,a_j$, it follows the vectors generated by the recursion method algorithm must be of the form
\begin{equation}
\begin{aligned}
  f_{2k} &= \sum_j D_j\,P_{k,j} (a_j + a^\dagger_j)\,, \,\,k=0,1,\dots K-1\quad  \quad \text{with}\quad P_{0,j}=1\,\,\forall\,j \\
  f_{2k+1} &= \sum_j D_j\,Q_{k,j}\,i (a^\dagger_j-a_j)\,, \,\,k=0,1,\dots K-1 \quad  \quad \text{with}\quad Q_{0,j}=\omega_j \,\,\forall\,j 
\end{aligned}
\label{eq:vec-ansatz}
\end{equation}
for some unknown real coefficients $P_{k,j}$ and $Q_{k,j}$ satisfying the above initial conditions. Using the harmonic oscillator thermal correlators
\begin{equation}
  \langle a_i\,a^\dagger_j\rangle_\beta = e^{\beta\omega_i}\,n_\beta(\omega_i)\,\delta_{ij}\,, \quad \quad \langle a^\dagger_i\,a_j\rangle_\beta = n_\beta(\omega_i)\,\delta_{ij},
\end{equation}
where $n_\beta(\omega) = (e^{\beta\omega}-1)^{-1}$, it follows
\begin{equation}
  \left(a_j + a_j^\dagger,\,a_k + a^\dagger_k\right) = \left(i(a_j^\dagger-a_j), i(a_k^\dagger - a_k)\right) = \frac{2}{\beta}\int^\beta_0 d\lambda\,e^{\lambda\omega_j}\,n_\beta(\omega_j) \delta_{jk} \equiv A_j\delta_{jk}\,,
\label{eq:norm-id}
\end{equation}
with all other inner product combinations vanishing. Hence, the set of vectors $\{f_{2k},\,f_{2k+1}\}$ defines an orthogonal set, as it should. For $K$ finite oscillators, the algorithm will halt once we reach a basis of $2K$ orthogonal vectors, which matches the number of independent creation/annihilation operators. This explicitly confirms the relation between the two set of operators is indeed simply a change of basis. This change is only non-trivial for $K > 1$, as in large-N QFTs where there is an infinite set of operator modes when studying the growth of a local field.

Plugging the parameterisation \eqref{eq:vec-ansatz} into \eqref{eq:rec-basic}, we obtain the following recurrence relations
\begin{equation}
\begin{aligned}
  P_{s,j} &= -\omega_j\,Q_{s-1,j} + \Delta_{2s-1}\,P_{s-1,j}\,, \quad \quad s=1,2,\dots K-1\\
  Q_{s,j} &= \omega_j\,P_{s,j} + \Delta_{2s}\,Q_{s-1,j}\,, \quad \quad s=1,2,\dots K-1
\end{aligned}
\label{eq:rec-rel}
\end{equation}
These are solved by\footnote{It is understood that whenever the subindex labels $r$ in $i_r$ equal zero, such terms do \emph{not} contribute to the solution.}
\begin{equation}
\begin{aligned}
  P_{s,j} &= \sum_{m=0}^s (-1)^m\,\omega_j^{2m}\,\sum_{i_1=1}^{2m+1} \Delta_{i_1} \sum_{i_2=i_1+2}^{2m+3} \Delta_{i_2}\dots \sum_{i_{s-m}=i_{s-m-1}+2}^{2s-1} \Delta_{i_{s-m}}\,, \\
  Q_{s,j} &= \omega_j \sum_{m=0}^s (-1)^m\,\omega_j^{2m}\,\sum_{i_1=1}^{2(m+1)} \Delta_{i_1} \sum_{i_2=i_1+2}^{2(m+2)} \Delta_{i_2}\dots \sum_{i_{s-m}=i_{s-m-1}+2}^{2s} \Delta_{i_{s-m}}\,.
\end{aligned}
\label{eq:rec-sol}
\end{equation}
The proof is by induction and it is given in App~(\ref{App5}). 

As a check of our formal solution to the recursion method, we show it satisfies the diffusion equation \eqref{eq:diffusion}. As stressed above, the advantage of expressing our operators in the basis of field modes is that these modes diagonalize the Liouvillian operator $\mathcal{L}$ and therefore make the time dependence trivial. Hence, the exact time evolution of our initial operator just equals
\begin{equation}
  f_0(t) = \sum_j D_j\,\left(e^{it\omega_j}a^\dagger_j + e^{-it\omega_j}\,a_j\right)\,.
\end{equation}
From this expression, to find the Lanczos expansion we just need to write the $a^\dagger_j$ and $a_j$ in terms of the $f_{k}$. We remark  this is not a dynamical question, just a change of basis. To invert the relation we use that the amplitudes entering the diffusion equation are given by the projections
\begin{equation}
  C_{2k}(t) \equiv \frac{(f_{2k},\,f_0(t))}{(f_{2k},\,f_{2k})}\,, \quad\quad C_{2k+1}(t) \equiv \frac{(f_{2k+1},\,f_0(t))}{(f_{2k+1},\,f_{2k+1})}\,.
\end{equation}
Explicit calculation yields
\begin{equation}\label{solution}
\begin{aligned}
  C_{2k}(t) &= \sum_j D_j^2\,A_j\,\frac{P_{k,j}}{(f_{2k},\,f_{2k})}\,\cos\omega_jt\,, \\
  C_{2k+1}(t) &= \sum_j D_j^2\,A_j\,\frac{Q_{k,j}}{(f_{2k+1},\,f_{2k+1})}\,\sin\omega_jt\,.
\end{aligned}
\end{equation}
Computing the time derivatives and using the recursion relations \eqref{eq:rec-rel} to replace the $\omega_j$ dependent terms reproduces the diffusion equation \eqref{eq:diffusion}.

To sum up, the 2-pt function determines the Lanczos coefficients $\Delta_{k}$ and these determine the orthonormal basis of operators $f_k$ controlling the Heisenberg time evolution of an initial operator $\mathcal{O}$. These are related by a change of basis to the operators used in section~\ref{sec:structure} to describe this same evolution in large-N gauge theories. This is consistent with our claim that 2-pt functions characterize operator growth in these theories at leading order. In this approximation, what makes these theories special, from the recursion method perspective, is that we can solve the diffusion equation analytically by relations \eqref{solution}, once we have the coefficients $\Delta_{k}$. In QFT, the set of modes is infinite and the recursion does not halt. The operator then grows indefinitely, even if we are dealing with a set of free harmonic oscillators.

The specific structure of the Lanczos basis unraveled here also differs from the general exponential growth of the Lanczos complexity in QFT. As remarked in \cite{Parker:2018yvk}, in the typical QFT scenario in which the 2-pt function is exponentially decaying and its Fourier transform has poles, as dictated by the generic analyticity properties of thermal correlations, then the mean position in the one-dimensional diffusion equation \eqref{eq:diffusion}, the Lanczos operator complexity \eqref{lc}, will grow exponentially fast with Lyapunov exponent $2\pi/\beta$.\footnote{As shown in appendix~\ref{inner-products}, this exponential growth with the right Lyapunov exponent only applies to the representative inner product considered in \eqref{innerL}. For other inner products, one gets exponential growths with faster rates. This statement seems similar to the results found in \cite{Romero-Bermudez:2019vej} for out-of-time-ordered correlation functions, which show that different choices of euclidean separations in the OTOC 4-pt function might lead to faster growth than the chaos bound \cite{Maldacena:2015waa}.} Such statement holds for any chaotic QFT, and it is not particular to large-N theories.


\subsection{Chaos and quantum complexity}
\label{sec:chaos}

In this section we stress the natural relation between operator growth and quantum circuit complexity, besides their connection through quantum chaos  \cite{Magan:2018nmu,Bueno2019,Barbon:2019tuq}. Moreover, in the context of the AdS/CFT correspondence, we also comment on the relation between them and the emergence of classical bulk chaos. 

In quantum complexity discussions, the complexity $\mathcal{C}_{\ket{\psi}}$ to prepare a particular target state $\ket{\psi}$ starting with a certain reference state $| \psi_\mt{R} \rangle$ by applying a series of elementary gates $g_i$ 
\begin{equation}\label{circuit}
	\ket{\psi}= U_\mt{R}\, \ket{\psi_\mt{R}}= g_{n}\cdots g_{2}\,g_{1}\ket{\psi_\mt{R}}\,,
\end{equation} 
is defined as the number of gates associated to the optimal protocol. Nielsen and collaborators \cite{2005quant.ph..2070N,2006Sci...311.1133N,2007quant.ph..1004D} mapped the problem of identifying this optimal circuit to the geometric problem of finding a geodesic in the space of unitaries acting on the Hilbert space. Given a one parameter family of states, labelled by $s$, the local driving hamiltonian $H(s)$ satisfies
\begin{equation}
  i \frac{d}{ds}\ket{\psi(s)}= H(s)\ket{\psi(s)}
\end{equation}
and generates the unitary transformation acting on the state
\begin{equation}
  U(\sigma) = \cev{\mathcal{P}} \exp \left[ -i \int^\sigma_0\!\!\! d s\, H(s)\right], \quad \text{with} \quad H(s)\equiv \sum_\text{I} Y^\text{I}(s)\,\mathcal{O}_\text{I}\,,
\end{equation}
where the Hermitian operators $\mathcal{O}_I$ generate the individual gates $g_\text{I}$. Circuits satisfying eq.~\eqref{circuit} correspond to trajectories satisfying the boundary conditions
\begin{equation}
  U(\sigma=0)= \mathbbm{1}\,, \qquad  U(\sigma=1)= U_\mt{R}\,.
\end{equation}
Optimal circuits minimise the cost defined as
\begin{equation}
  \mathcal{C}_{\ket{\psi_\mt{R}}\rightarrow\ket{\psi}} \equiv \int^1_0 ds ~ F \left( H(s)\right)
\end{equation}
where $F$ is a local cost function depending on the tangent vector $H(s)$.

Consider two states, $\ket{\psi}$, as above, and a perturbed state $\ket{\psi_{\mathcal{O}}}=e^{i\mathcal{O}}\ket{\psi}$, generated by the action of a simple unitary generated by certain local operator $\mathcal{O}$. The time evolution of both states is determined by the unitary action $U(t)=e^{-iHt}$ on them, where $H$ corresponds to the physical hamiltonian of the system, leading to the states $\ket{\psi (t)}$ and $\ket{\psi_{\mathcal{O}}(t)}$, respectively. In this set-up, one can define some notion of growth or size based on the circuit complexity to go from one evolved state to the other\footnote{This might be related to the complexity variation $\mathcal{C}_{\ket{\psi (t)}} - \mathcal{C}_{\ket{\psi_{\mathcal{O} }(t)}}$ considered in \cite{Barbon:2019tuq}. Variations in quantum circuit complexity have also been considered recently in \cite{Bueno2019,Bernamonti:2019zyy}.}
\begin{equation}
S_{\mathcal{O}(t)}\equiv   \mathcal{C}_{\ket{\psi (t)}\rightarrow\ket{\psi_{\mathcal{O} }(t)} }
\end{equation}

This relative complexity is simpler than expected in the limit of small perturbations. As shown in the original geometric complexity paper \cite{2005quant.ph..2070N}, if the perturbation is small enough so that  $\ket{\psi}$ and $\ket{\psi_{\mathcal{O}}}$ are sufficiently closed to each other, the geodesic connecting them is simply
\begin{equation}
U(s)_{\ket{\psi}\rightarrow\ket{\psi_{\mathcal{O}}}}= e^{i\mathcal{O}s}\,, \qquad 0\leqslant s\leqslant 1
\end{equation}
The key observation now is that the geodesic connecting the time evolved states $\ket{\psi (t)}$ and $\ket{\psi_{\mathcal{O} (t)}}$ is also going to be of the same type
\begin{equation}
U(s)_{\ket{\psi (t)}\rightarrow\ket{\psi_{\mathcal{O}}(t)} }= e^{i\mathcal{O}(-t) s}\,, \qquad 0\leqslant s\leqslant 1\,,
\end{equation}
by dialling the initial perturbation to be small enough. This argument allows to write the relative complexity of such geodesic as
 \begin{equation}\label{circuitt}
S_{\mathcal{O}(t)}\equiv   \mathcal{C}_{\ket{\psi (t)}\rightarrow\ket{\psi_{\mathcal{O} }(t)} }= \int^1_0 ds ~ F \left( \mathcal{O}(-t)\right) = F \left( \mathcal{O}(-t)\right)\;.
\end{equation}
The precise evaluation requires a choice of the cost function. See \cite{Bueno2019} for a discussion and calculation of several possibilities.

\paragraph{Connection to previous notions of size.} Equation \eqref{circuitt} states that the circuit complexity is the computational cost of the time evolved operator responsible for the perturbation.\footnote{The connection between cost functions and operator size was recognized already in \cite{Magan:2018nmu} for spin systems, but equation~\eqref{circuitt} shows it holds more generally.}  The final value depends on the choice of a cost function, pretty much as in our earlier discussions on operator size, the latter depends on the definition of size. Crucially, \eqref{circuitt} stresses that, given some cost function, circuit complexity only depends on the time evolution of the operator, the same structure controlling any notion of operator growth or size.
Hence, both notions are functionally dependent. In particular, if we were to define the cost as one of the previous notions of operator size, both would be equivalent. A convenient choice then is the Minkowski energy discussed in the previous section. With this choice, the relative complexity of large-N theories will grow exponentially fast with Lyapunov exponent $\lambda= 2\pi/\beta$. \footnote{This construction and the Minkowski energy choice for the complexity cost provides a specific realization of the idea put forward in \cite{Magan:2018nmu}, in which the cost function was argued to be related to the scaling dimension of the associated perturbation.}

\paragraph{Connection to chaos.} Chaotic behavior concerns the sensitivity of certain dynamical systems to small perturbations $\delta x_{i} $ of its initial conditions. Classically, such sensitivity is usually studied in a double scaling limit, where the size of the perturbation is taken to zero first and the limit of large times is taken afterward. The first limit ensures that a linearized equation of the type
\begin{equation}\label{chaosclas}
\delta x_{i} (t)=\sum\limits_{j}M_{ij}\delta x_{i}\;,
\end{equation}
is a good approximation to the dynamics, where the Jacobian matrix $M_{ij}=\partial x_{i}(t)/\partial x_{j}$ encodes the dependence on the initial conditions $x_j$. The second limit ensures the solution to \eqref{chaosclas} is dominated by the largest Lyapunov exponent.

A natural extension of the classical definition to the quantum domain has been recently developed in \cite{Bueno2019}. As shown in \cite{Ashtekar:1997ud}, quantum dynamics can be formulated as classical dynamics on a ``quantum phase space'', defined to be the Hilbert space itself. The symplectic form at point $\vert\psi\rangle$ along two infinitesimal directions, generated by Hamiltonian operators $H_{1}$ and $H_{2}$ is just the expectation value of the commutator $\Omega_{\vert\psi\rangle}(H_{1},H_{2})\equiv\langle\psi\vert [H_{1},H_{2}]\vert\psi\rangle$. It turns out that Schr\"odinger equations are seen as Hamilton equations in such a phase space with ``classical Hamiltonian'' $H(\vert\psi\rangle)=\langle\psi\vert \hat{H}\vert \psi\rangle$. Having framed quantum dynamics as a classical system, it is most natural to define quantum chaos by the usual classical definition~(\ref{chaosclas}) but applied to the quantum phase space. This definition has, by construction, the appropriate pullback to the classical definition on a semiclassical phase space, but it is otherwise valid through the whole quantum system.

The classical approach to quantum mechanics illuminates the relation between operator growth and chaos by showing the transparent relation between $\mathcal{O}(-t)$, the operator that generates the unitary interpolating between the nearby quantum states at time $t$, and the Jacobian matrix associated to the classical chaotic process. In the Hilbert space we can define generalized coordinates $\vert q_{i},p_{i}\rangle$ (at least locally) satisfying the canonical Poisson brackets with respect to the Hilbert space symplectic form. Generic infinitesimal perturbations of any state can be written as\footnote{There is a missing phase in the equation, due to the non-commutativity between $q$ and $p$. It is not included here because it is second order
in the infinitesimal perturbations $\delta q ,\delta p$.}
\begin{equation}
e^{i\mathcal{O}}\vert q_{i},p_{i}\rangle \equiv e^{i(\hat{p}_{i}\delta q_{i}-\hat{q}_{i}\delta p_{i})}\vert q_{i},p_{i}\rangle =\vert q_{i} +\delta q_{i},p_{i}+\delta p_{i}\rangle \,\,\,\,\,\Longrightarrow\,\,\,\,\, \mathcal{O}=\sum\limits_{i}\hat{p}_{i}\delta q_{i}-\hat{q}_{i}\delta p_{i}\;.
\end{equation}
This equation just states that any small perturbation $\mathcal{O}$ can be expanded in the generators of translations along the local reference frame defined by $q_{i},p_{i}$. Evolving in time, one 
observes
\begin{equation}
\mathcal{O} (-t)=\sum\limits_{i}\hat{p}_{i}(-t)\delta q_{i}-\hat{q}_{i}(-t)\delta p_{i}=\sum\limits_{i}\hat{p}_{i}\delta q_{i} (t)-\hat{q}_{i}\delta p_{i} (t)\;,
\end{equation}
where $\delta q_{i} (t)$ and $\delta p_{i} (t)$ are determined by a linearized equation of the type~(\ref{chaosclas}) associated to the quantum phase space. This is analyzed in a specific generic example in \cite{Bueno2019}. It follows that the growth properties of the perturbation $\mathcal{O}$ are controlled by the Jacobian matrix defining the chaotic process in the quantum phase space, and vice versa.

\paragraph{Classical chaos in AdS/CFT.}
In the context of the AdS/CFT correspondence, CFT perturbations $e^{i\mathcal{O}}\ket{\psi}$ generated by operators $\mathcal{O}$ with large conformal dimension can be described by freely falling particles in the bulk geometry dual to $\ket{\psi}$, in the semiclassical approximation \cite{Fitzpatrick:2011jn,Nozaki:2013wia,Goto:2016wme,Goto:2017olq,Terashima:2017gmc,Berenstein:2019tcs}. At high energies, the bulk geometry involves a black hole with the temperature related to the energy of the state by the usual thermodynamic relation.

It was realized in \cite{Barbon:2011pn,Barbon:2011nj,Barbon:2012zv} that the optical metric capturing the near horizon physics has chaotic behavior, i.e. the optical metric is controlled by a hyperbolic space whose geodesics are known to have chaotic properties. The hyperbolic space (universally attached to any horizon) turns out to have a radius of curvature given by $R=\beta/2\pi$, and therefore the associated bulk geodesic deviation growth is compatible with the holographic Lyapunov exponent. Such deviation rests on the conformal transformation from the near horizon Rindler geometry to the hyperbolic one. It therefore secretly rests on the emergent Poincar\'e symmetries described above. The fact that Minkowski energies and radial momenta grow exponentially in the Rindler frame with Lyapunov exponent $\lambda=2\pi/\beta$ is mapped, in the optical frame, to the fact that perturbations in the transverse direction grow exponentially fast with the same Lyapunov exponent. The conformal transformation from the Rindler frame to the optical one was studied at the classical level and also in the QFT setup in \cite{Barbon:2012zv}. Having constructed the Poincar\'e symmetries above from the structure of time evolution, such conformal transformation to the optical frame can be constructed as well, as if we were in the Minkowski/Rindler scenario, and the chaotic properties of the optical metric are thus recovered.


\section{Discussion}
\label{Discussion}

We have considered the problem of operator growth in large-N gauge theories at finite temperature. We framed the problem as that of understanding the operators
\begin{equation}
\mathcal{O}_{n}  \equiv [H,\cdots,[H,\mathcal{O}]\cdots]
\end{equation}
where $H$ is the Hamiltonian. This is most natural since the expansion of Heisenberg time evolution in powers of time \eqref{Heis} teaches us these are the only operators with whom the initial operator mixes over time. 

In section~\ref{sec:structure}, we argued that the expression
\begin{equation}\label{no2}
\mathcal{O}_{n}(t)=\int_{\omega > 0} \frac{d\omega d^{d-1}\vec{k}}{(2\pi)^d}\, \left((-\omega)^{n}\,\mathcal{O}_{\omega,\vec{k}}\,e^{-i\omega t + i\vec{k}\vec{x}} + \,\omega^{n}\,\mathcal{O}^\dagger_{\omega,\vec{k}}\,e^{i\omega t - i\vec{k}\vec{x}}\right)\,
\end{equation}
together with linearity, large-N factorization, the fact that most eigenstates at a given temperature behave as thermal states (ETH) and the correlation functions of the field modes $\mathcal{O}_{\omega,\vec{k}}$ and $\mathcal{O}^\dagger_{\omega,\vec{k}}$, given by \eqref{eq:exp}, completely specify the action of these operators in most interesting states, up to $1/N$ corrections. Hence, \eqref{no2} determines the time evolution of the operator $\mathcal{O}(t,\vec{x})$. A similar statement holds for modular time evolution as well. 

A first interesting insight is that \emph{any} notion of operator growth should be determined by the 2-pt function at this order. Given the relation found between operator growth and four-point functions at an infinite temperature in SYK \cite{Roberts:2018mnp}, one might hope for a non-trivial relation between the two-point function and the connected four-point function in generic large-N theories. We leave this interesting observation for future work.

In section~\ref{sec:algebra}, we constructed an emergent \emph{bulk} Poincar\'e algebra \eqref{pal}
as the first application of our proposed solution. This was achieved by the known doubling of the modes appearing in large-N theories \cite{Papadodimas:2012aq}. This algebra is related to the near horizon Rindler behaviour of thermal horizons. Albeit we have focused on the conventional modes, defined by means of the Hamiltonian of the large-N theory, the construction can be easily extended to modular time evolution, by using the modular modes~(\ref{modular}) instead of the conventional ones. If large-N factorization holds, we can again find renormalized modes satisfying the algebra of free creation and annihilation operators, and proceed with the construction of an emergent Poincar\'e algebra. It would be interesting to develop the arguments given here in relation to the recent results in \cite{deBoer:2019uem}, based on prior work \cite{Czech:2019vih}, where such Poincar\'e algebra emerges from a local bulk perspective due to the limiting modular evolution behaviour, making the latter much closer to Equivalence Principle considerations.

Albeit the simplicity of the previous solution contrasts with the expected complexity of the problem, in section~\ref{sec:growth}, we analyzed several existent notions of operator growth and size existent in the literature from the perspective presented here. These include number operators, energy measures, the recursion method in condensed matter physics, and the approach to quantum chaos based on quantum circuit complexity. All of them can be considered in detail, and analytically, in the basis of field modes. We have seen that the different approaches are just variations over a common theme: the evolution in time of the initial operator $\mathcal{O}$, which is fully characterized by expression \eqref{exparep}, at leading order. 
We made proposals for operator size in large-N QFTs by noticing that the GNS construction maps operator evolution to conventional state evolution in the GNS Hilbert space, and also derived an explicit relation between operator complexity and circuit complexity in eq~\eqref{circuitt}. In the large-N limit, all such notions are functionals of the two-point function alone.\footnote{The dependence of operator growth on the two-point function was noticed in Ref. \cite{Magan:2016ehs}, for the case of the operator growth of a ``complex'' operator in the vacuum.} 

It is an important open problem to understand how to systematically incorporate $1/N$ corrections to our discussion. Given our approach, this is not an intrinsic problem attached to operator growth, but it is generic to large-N gauge theories including holographic ones if one is interested in a bulk interpretation of these statements. See \cite{Mousatov:2019xmc} for a recent discussion on how to incorporate these corrections in the bulk for some choices of operator size. 

We finish by stressing a point made in the introduction: our work neatly shows that \emph{any} notion of operator growth has an equivalent formulation both in the bulk and the boundary theories, in the context of large-N holographic theories. This is just a consequence of the equality between bulk and boundary Hilbert space and Hamiltonians. In our setup, on a technical level, this is transparent due to the work in bulk reconstruction relating bulk field modes with boundary modes $\mathcal{O}_{\omega,\vec{k}}$, $\mathcal{O}^\dagger_{\omega,\vec{k}}$, and their mirror partners, mainly following \cite{Papadodimas:2012aq,Papadodimas:2013wnh,Papadodimas:2013jku}.


\section*{Acknowledgments}

\noindent We thank Jos\'e Barb\'on, Pablo Bueno, Horacio Casini and Simon Ross for useful discussions. We also want to thank the Kavli Institute for Theoretical Physics and the Higgs Centre for Theoretical Physics for hospitality and financial support during the workshops "Chaos and Order" and "Recent development in holography", respectively. The work of JMM was supported by the Simons foundation through the It From Qubit Simons collaboration.  

\appendix

\section{The Gelfand-Naimark-Segal (GNS) construction}
\label{App2}

The GNS construction \cite{GelNeu43,segal1947} generates a Hilbert space $\mathcal{H}_{\omega}$ from an abstract $C^\star$-algebra $\mathcal{A}$ and a linear functional (state) $\omega$ from $\mathcal{A}$ to $\mathbb{C}$, together with a representation of the algebra $\pi (\mathcal{A})$ acting on it. In this appendix, we review its main ideas following closely  \cite{Haag:1992hx}.

A $C^\star$-algebra $\mathcal{A}$ is a set of objects such that if $A,B\in \mathcal{A}$ and $a,b\in \mathbb{C}$, then the linear combination $aA+bB\in \mathcal{A}$. Furthermore, there exists a map $A\mapsto A^{\dagger}\,,$ $\forall\,A\in\mathcal{A}$ being an involution and satisfying 
\begin{equation}
  (AB)^\dagger=B^\dagger A^\dagger\,, \quad (aA)^\dagger = a^\star\,A^\dagger \quad \quad \forall A,B \in \mathcal{A}\,, \quad \forall a\in\mathbb{C}
\end{equation}
Up to topological requirements, see \cite{Haag:1992hx} for a more detailed account, $\mathcal{A}$ is called a von Neumann algebra if it further contains the identity. From now on we consider von Neumann algebras only.

States $\omega$ are positive and normalized linear functionals from $\mathcal{A}$ to $\mathbb{C}$ satisfying
\begin{equation}
\begin{aligned}
  \omega (aA+bB)&= a\,\omega(A)+b\,\omega (B)\,, \\
  \omega (A^\star A)&\geq 0\,,\\
  \omega (\mathds{1})&=1\,.
\end{aligned}
\label{eq:omega-state}
\end{equation}

Hilbert spaces $\mathcal{H}$ are vector spaces with an inner product mapping any pair of elements $\vert w\rangle,\vert v\rangle\in\mathcal{H}$ to a complex number $\langle v\vert w\rangle \in \mathbb{C}$ satisfying
\begin{equation}
\begin{aligned}
  \langle v\vert w\rangle &=\langle w\vert v\rangle^\star\,, \\
  \langle av_{1}+bv_{2}\vert w\rangle &=a^\star\langle v_{1}\vert w\rangle+b^\star \langle v_{2}\vert w\rangle\,, \\
  \vert\langle v\vert v\rangle\vert^{2}&\geq  0\,,
\end{aligned}
\label{eq:inner-def}
\end{equation}
where the last line is only saturated for $ \vert v\rangle =0$. 

Let the algebra $\mathcal{A}$ be an algebra of operators. Since the algebra $\mathcal{A}$ is already a vector space over $\mathbb{C}$, to become a Hilbert space it requires an inner product. This can be defined using the state $\omega$ as
\begin{equation}
  \langle A\vert B\rangle = \omega (A^{\dagger}B)\,,
\label{eq:inner-map}
\end{equation}
where we used the standard notation in quantum mechanics $|A\rangle$ to refer to the vector in $\mathcal{H}$ associated with the operator $A\in \mathcal{A}$.

This inner product satisfies all the requirements \eqref{eq:inner-def} except for the existence of non-zero operators $W$ satisfying $\omega (W^{\dagger}W)=0$. The set of such operators $\mathcal{I}$ is a left ideal in $\mathcal{A}$, the so called Gelfand ideal of the state $\omega$, i.e a linear subspace of $\mathcal{A}$ that is stable under multiplication by any element $A\in \mathcal{A}$ from the left
\begin{equation}
W\in \mathcal{I}\,\, ,\,\,\, A\in \mathcal{A}\Rightarrow AW\in\mathcal{I}\,.
\end{equation}
The GNS construction defines the Hilbert space $\mathcal{H}_\omega$ as the quotient of $\mathcal{H}$ by the ideal $\mathcal{I}$, i.e. $\mathcal{H}_{\omega}\equiv \mathcal{A}/\mathcal{I}$ \footnote{More precisely $\mathcal{H}_{\omega}$ is defined as the completion of $\mathcal{A}/\mathcal{I}$ with respect to the norm topology.}. Vectors $\vert [A]\rangle \in \mathcal{H}_{\omega}$ correspond to equivalence classes of operators in the algebra of the form $A+\mathcal{I}$ and such classes do not depend on the representative.

GNS induces a representation $\pi_{\omega}$ of $\mathcal{A}$ acting on $\mathcal{H}_{\omega}$ by the product in the algebra $\mathcal{A}$
\begin{equation}
  \pi_{\omega}(A)\vert [B]\rangle =\vert [AB]\rangle\;.
\label{pi-repr}
\end{equation}
A consequence of this construction is that the identity class $\vert \Omega\rangle\equiv\vert [\mathds{1}]\rangle$ vector can be associated to the starting state $\omega$ since
\begin{equation}
  \omega (A)=\langle \Omega\vert A \vert \Omega\rangle\,.
\end{equation}

\subsection{GNS of the thermal state}
\label{sec:GNS-thermal}

When the GNS construction is considered for finite-dimensional algebras $\mathcal{A}$ containing bounded operators, the identity operator $\mathds{1}$ can be understood as the maximally entangled density matrix (up to normalization) and the inner product \eqref{eq:inner-map} can be taken as
\begin{equation}
  \langle A|B\rangle = \frac{1}{Z}\text{Tr}\left(A^\dagger B\right)\,,
\label{eq:infinite}  
\end{equation}
where $Z\equiv \text{Tr}(\mathds{1})$. There is no Gelfand ideal $\mathcal{I}$ in this case. Hence, there exists an isomorphism between $\mathcal{A}$ and $\mathcal{H}_\omega$. The same conclusion holds when one replaces $\mathds{1}$ with $\kappa = \rho_\beta^{1/2}$, where $\rho_\beta$ is the Boltzmann finite temperature density matrix (or any density matrix of full rank). 

Besides the GNS representation $\pi(\mathcal{A})$
\begin{equation}
  \pi (A)\vert \kappa\rangle \equiv\vert A\kappa\rangle\,,
\end{equation}
there exists the conjugate representation $\bar{\pi}(\mathcal{A})$, defined by
\begin{equation}
  \bar{\pi}(A)\vert \kappa\rangle \equiv\vert \kappa A^{\dagger}\rangle\,.
\end{equation}
These are equivalent because there exists an anti-unitary operator $J$ acting on $\mathcal{H}_{\omega}$ satisfying
\begin{equation}
  J\vert A\kappa\rangle = \vert \kappa A^{\dagger}\rangle \,\,\,\,\,\,\text{with} \,\,\,\,\,\,\, J^{2}=1\,,
\end{equation}
implying
\begin{equation}
  J\pi (A)J= \bar{\pi}(A)\,.
\end{equation}
It also follows from these expressions that
\begin{equation}
  \omega (A)=\langle\kappa\vert\pi (A)\vert \kappa\rangle =\langle\kappa\vert \bar{\pi} (A)\vert \kappa\rangle^{*}\,.
\label{eq:state-beta}
\end{equation}
Since $\vert\kappa\rangle$ is invariant under time evolution\footnote{For a general full rank state $\rho$, the evolution which is naturally defined by this construction is the so-called modular evolution. If $\rho= e^{-H}$, with $H$ the modular Hamiltonian, then $H$ is the generator of modular time evolution.}, we can now easily define unitary evolution $U_{t}$ in all states of the representation by
\begin{equation}\label{ugns1}
U_{t}\pi (A)\vert \kappa\rangle =\pi (A_{t})\vert \kappa\rangle \,, \,\,\,\,\,\,\,\,\,\,\,\,\, U_{t}\bar{\pi} (A)\vert \kappa\rangle =\bar{\pi} (A_{t})\vert \kappa\rangle\;.
\end{equation}
It helps to disentangle the meaning of these definitions to explicitly write the unitary evolution as
\begin{equation}\label{ugns2}
U_{t}=\pi (e^{iHt})\bar{\pi} (e^{iHt})\;.
\end{equation}
We can then check the definition \eqref{ugns1}
\begin{equation}
U_{t}\pi (A)\vert \kappa\rangle =U_{t}\vert A\kappa\rangle= \vert e^{iHt}A\kappa e^{-iHt}\rangle=\vert e^{iHt}Ae^{-iHt}\kappa \rangle=\vert A_{t}\kappa \rangle=\pi (A_{t})\vert \kappa\rangle
\end{equation}
is satisfied. Given this representation, it is natural to introduce the full hamiltonian as $H_{\textrm{F}}=\pi (H)-\bar{\pi} (H)$.
Using $\kappa = \rho_\beta^{1/2} = Z^{-1/2} e^{-\beta H/2}$, it follows
\begin{equation}
  J e^{-\beta H_{\textrm{F}}/2}\pi (A)\vert\kappa\rangle =J e^{-\beta H_{\textrm{F}}/2}\vert A\kappa\rangle =J \vert\kappa A\rangle =J \bar{\pi}(A^{\dagger})\vert\kappa \rangle = \pi (A^{\dagger})\vert \kappa\rangle \;.
\label{tomit}  
\end{equation}
It is interesting to single out the equality from the first term to the fourth term. Multiplying from the left both side by $J$ and moving the right hand side to the left we obtain
\begin{equation}
   (e^{-\beta H_{\textrm{F}}/2}\pi (A)-\bar{\pi}(A^{\dagger}))\vert\kappa\rangle =0\;,
\label{tomit2}  
\end{equation}
which is the (generalized) origin of the known relations~(\ref{trind}) and (\ref{tomitO}), associated to free QFT and large-N theories.

As stressed through the article, one general lesson is that operator evolution can be seen as a conventional state evolution through the GNS construction. The generator of the GNS unitary evolution, the GNS Hamiltonian, is what in the condensed matter community is called the Liouvillian \cite{viswanath2008recursion}, see \cite{Parker:2018yvk,Barbon:2019wsy} for recent applications of such approach. Furthermore, the GNS construction points out the subtlety of the Gelfand ideal, stressing why states with full rank are convenient, and allows a direct application into QFT since it is valid for all types of algebras.


\subsection{Inner products in the space of operators}
\label{inner-products}

In the previous GNS construction, one starts with a natural inner product on the space of operators, such as \eqref{eq:infinite} or \eqref{eq:state-beta}. This choice of inner product is not unique. Denoting a general inner product by 
$(A,B)$, in \cite{viswanath2008recursion} the following family is considered
\begin{equation}\label{inner}
(A,B)=\frac{1}{\beta}\int\limits_{0}^{\beta}\,d\lambda\,g(\lambda)\, \langle e^{\lambda H}A^{\dagger}e^{-\lambda H} B\rangle_{\beta}-\langle A^{\dagger}\rangle_\beta \langle B\rangle_\beta\;,
\end{equation}
where $\langle A \rangle_\beta \equiv \text{Tr}(\rho_\beta\,A)$ and $g(\lambda)$ is any function satisfying:
\begin{equation}
g(\lambda)\geq 0 \,\,\,\,\,\,\,\, g(\beta-\lambda)=g(\lambda)\,\,\,\,\,\,\,\,\,\,\frac{1}{\beta}\int\limits_{0}^{\beta}d\lambda\, g(\lambda)=1\;.
\end{equation}
Examples of such functions are
\begin{equation}
g(\lambda)=\frac{1}{2}\beta\, [\,\delta (\lambda)+\delta (\beta-\lambda)]\, , \,\,\,\,\,\,\,\,\, g(\lambda)=\delta (\beta/2-\lambda)\;,
\end{equation}
for which the inner product reduces to
\begin{equation}
(A,B)=\frac{1}{2}\langle A^{\dagger} B+B A^{\dagger}\rangle-\langle A^{\dagger}\rangle \langle B\rangle\, , \,\,\,\,\,\,\,\,\, (A,B)= \langle e^{\beta H/2}A^{\dagger}e^{-\beta H/2} B\rangle_{\beta}-\langle A^{\dagger}\rangle \langle B\rangle\;,
\end{equation}
respectively. In this section we want to describe the status of this big family~(\ref{inner}) of inner products. In particular, we show below how they change the specific functional describing the chaotic growth. Indeed faster growths than the chaos bound can be obtained, albeit there is no surprise here, since one can actually relate all the inner products~(\ref{inner}) to the one defined by $g(\lambda)=\beta\delta (\beta/2-\lambda)$, as we show below.

To test the dependence on the inner product we can analyze the basic quantity controlling the growth, which is the return probability
\begin{equation}
p(t)\equiv\vert ( \mathcal{O}(t)\vert \mathcal{O}(0) )\vert^{2}\;.
\end{equation}
Using the previous inner products we thus need to analyze\footnote{We assume the operator $\mathcal{O}$ to have vanishing one point function for visual clarity. Including in the discussion one-point functions is trivial since they do not depend on time.}
\begin{equation}
(\mathcal{O}(t),\mathcal{O}(0))=\frac{1}{\beta}\int\limits_{0}^{\beta}\,d\lambda\,g(\lambda)\, \langle e^{\lambda H}\mathcal{O}(t)e^{-\lambda H} \mathcal{O}\rangle_{\beta}=\frac{1}{\beta}\int\limits_{0}^{\beta}\,d\lambda\,g(\lambda)\, \langle \mathcal{O}(t-i\lambda) \mathcal{O}\rangle_{\beta}\;,
\end{equation}
and such function is completely determined equivalently by its Fourier transform
\begin{equation}
R(\omega)=\int e^{i\omega t}\,(\mathcal{O}(t),\mathcal{O}(0))\;.
\end{equation}
To find such Fourier transform first consider:
\begin{equation}\label{gl}
G_{\lambda}(\omega)\equiv \int e^{i\omega t}\,\langle \mathcal{O}(t-i\lambda) \,\mathcal{O}\rangle_{\beta}\;.
\end{equation}
The expectation value $\langle \mathcal{O}(t) \,\mathcal{O}\rangle_{\beta}$ can be analytically continued to imaginary times for $0> \textrm{Im} (t)> -\beta$. Call $F(z)$ such a unique function in the strip. We can consider the following integrals of such function:
\begin{equation}
\int\limits_{\mathcal{C}_{\lambda}} F(z) e^{iz\omega}dz\;,
\end{equation}
where $\mathcal{C}_{\lambda}$ is a path parallel to the real time axis, shifted by $-i\lambda$. Due to the absence of poles or singularities in such a region, Cauchy's theorem implies:
\begin{equation}
\int\limits_{\mathcal{C}_{\lambda}} F(z) e^{iz\omega}dz = \int\limits_{\mathcal{C}_{\lambda'}} F(z) e^{iz\omega}dz \;.
\end{equation}
This implies that:
\begin{equation}
\int\limits_{-\infty}^{\infty} F(t-i\lambda)e^{i(t-i\lambda)\omega}dt=\int\limits_{-\infty}^{\infty} F(t-i\lambda')e^{i(t-i\lambda')\omega}\;.
\end{equation}
Looking at~(\ref{gl}), we conlude that
\begin{equation}
e^{\lambda\omega}G_{\lambda}(\omega)=e^{\lambda'\omega}G_{\lambda'}(\omega)\;.
\end{equation}
This relation is important, since it says that knowing the exact Fourier transform at some imaginary time $\lambda$ is equivalent to knowing it at all complexified times in the holomorphic strip. In particular, for 2d CFT's, we have the exact results for $\lambda=\beta/2$, so that for general $0<\lambda<\beta$:
\begin{equation}
G_{\lambda}(\omega)=e^{-\lambda\omega}e^{\beta\omega/2}G_{\beta/2}(\omega)\;.
\end{equation}
The Fourier transform of the autocorrelation function ends up being:
\begin{equation}
R(\omega)=e^{\beta\omega/2}G_{\beta/2}(\omega)\frac{1}{\beta}\int\limits_{0}^{\beta}\,d\lambda\,g(\lambda)\,e^{-\lambda\omega}\;,
\end{equation}
which is a simple functional of the information encoded in the inner product defined by $g(\lambda)=\beta\delta (\beta/2-\lambda)$.

The previous relation implies that all inner products have slower decay tails of $R(\omega)$ at large $\omega$ than the choice $g(\lambda)=\beta\delta (\beta/2-\lambda)$. The choice $g(\lambda)=\beta\delta (\beta/2-\lambda)$ is indeed the one considered in  \cite{Parker:2018yvk}, which in chaotic QFT leads to a exponential growth with Lyapunov exponent $\lambda=2\pi/\beta$. Therefore, we conclude that all other choices display stronger Lanczos growths than the one expected by the chaos bound. We point out the existence of the results found in \cite{Romero-Bermudez:2019vej}, in the context of OTOCs, which show that different choices of euclidean separations in the 4-pt functions lead to faster growths than the chaos bound \cite{Maldacena:2015waa}.

\section{Solving the recurrence relation}
\label{App5}

In this appendix we complete the proof of the solution of the Lanczos recursion method when applied to large-N theories. Given $P_{0,j}$ and $Q_{0,j}$ in \eqref{eq:vec-ansatz}, the ansatz \eqref{eq:rec-sol} reproduces correctly $P_{1,j}$ and $Q_{1,j}$.
Assuming \eqref{eq:rec-sol} holds for $s$, we want to show $P_{s+1,j}$ and $Q_{s+1,j}$ satisfy \eqref{eq:rec-sol}. Let us use the first equation in \eqref{eq:rec-rel} to compute $P_{s+1,j}$
\begin{equation}
\begin{aligned}
  P_{s+1,j} &= -\omega_j\,Q_{s,j} + \Delta_{2s+1}\,P_{s,j} \\
  &= \sum_{m=0}^s (-1)^{s+1}\omega_j^{2(m+1)} \,\sum_{i_1=1}^{2(m+1)} \Delta_{i_1} \sum_{i_2=i_1+2}^{2(m+2)} \Delta_{i_2}\dots \sum_{i_{s-m}=i_{s-m-1}+2}^{2s} \Delta_{i_{s-m}} \\
  & + \Delta_{2s+1}\,\sum_{r=0}^s (-1)^r\,\omega_j^{2r}\,\sum_{i_1=1}^{2r+1} \Delta_{i_1} \sum_{i_2=i_1+2}^{2r+3} \Delta_{i_2}\dots \sum_{i_{s-r}=i_{s-r-1}+2}^{2s-1} \Delta_{i_{s-r}} \\
  &= (-1)^{s+1}\omega_j^{s+1} + \sum_{r=1}^s (-1)^r\,\omega_j^{2r} \,\sum_{i_1=1}^{2r} \Delta_{i_1} \sum_{i_2=i_1+2}^{2(r+1)} \Delta_{i_2}\dots \sum_{i_{s-m}=i_{s+1-r}+2}^{2s} \Delta_{i_{s-m}} \\
  & + \Delta_{2s+1}\,\sum_{r=0}^s (-1)^r\,\omega_j^{2r}\,\sum_{i_1=1}^{2r+1} \Delta_{i_1} \sum_{i_2=i_1+2}^{2r+3} \Delta_{i_2}\dots \sum_{i_{s-r}=i_{s-r-1}+2}^{2s-1} \Delta_{i_{s-r}}\,.
\end{aligned}
\label{eq:rec-step}
\end{equation}
To derive the third equality, we relabelled $m+1=r$ and wrote the $r=s+1$ term separately, i.e. the only one involving no $\Delta$s. Notice that the only $\omega_j^0$ term comes from $r=0$ in the last line and reproduces the right answer $\Delta_{2s+1}\Delta_1\Delta_3\dots \Delta_{2s-1}$. Hence, given an index $r=\{1,2,\dots s\}$, the task is to show
\begin{equation}
  \sum_{i_1=1}^{2r+1} \Delta_{i_1} \sum_{i_2=i_1+2}^{2r+3} \Delta_{i_2}\dots \sum_{i_{s+1-r}=i_{s-r-1}+2}^{2s+1} \Delta_{i_{s+1-r}}
\label{eq:rec-final}
\end{equation}
equals the sum of the coefficients in the last two lines of \eqref{eq:rec-step} for a given $r$. The key point is the different upper limit in the last sum in \eqref{eq:rec-final} being $2s+1$. Indeed, when $i_{s+1-r}\neq 2s+1$, the contribution from \eqref{eq:rec-final} equals the first line in \eqref{eq:rec-step}. Hence, we are left to show the contribution from $i_{s+1-r} = 2s+1$ in \eqref{eq:rec-final}
\begin{equation}
  \Delta_{2s+1} \sum_{i_1=1}^{2r+1} \Delta_{i_1} \sum_{i_2=i_1+2}^{2r+3} \Delta_{i_2}\dots \sum_{i_{s-r}=i_{s-r-1}+2}^{2s-1} \Delta_{i_{s-r}}
\end{equation}
equals the last line in \eqref{eq:rec-step}, which it does. 

We are left to check $Q_{s+1,j}$ satisfies
\begin{equation}
  Q_{s+1,j} = \omega_j \sum_{m=0}^{s+1} (-1)^m\,\omega_j^{2m}\,\sum_{i_1=1}^{2(m+1)} \Delta_{i_1} \sum_{i_2=i_1+2}^{2(m+2)} \Delta_{i_2}\dots \sum_{i_{s+1-m}=i_{s-m}+2}^{2(s+1)} \Delta_{i_{s+1-m}}
\label{eq:rec-step2}
\end{equation}
having assumed $P_{s+1,j}$ and $Q_{s,j}$ do satisfy \eqref{eq:rec-sol}. Using the second recursion relation \eqref{eq:rec-sol}, we can write $Q_{s+1,j}$ as
\begin{equation}
\begin{aligned}
 &  \omega_j\,P_{s+1,j} + \Delta_{2(s+1)}\,Q_{s,j} = \\
 &\omega_j \sum_{m=0}^{s+1} (-1)^m\,\omega_j^{2m}\,\sum_{i_1=1}^{2m+1} \Delta_{i_1} \sum_{i_2=i_1+2}^{2m+3} \Delta_{i_2}\dots \sum_{i_{s+1-m}=i_{s-m}+2}^{2s+1} \Delta_{i_{s+1-m}}\\
 &+ \Delta_{2(s+1)}\,\omega_j \sum_{m=0}^{s} (-1)^m\,\omega_j^{2m}\,\sum_{i_1=1}^{2(m+1)} \Delta_{i_1} \sum_{i_2=i_1+2}^{2(m+2)} \Delta_{i_2}\dots \sum_{i_{s-m}=i_{s-m-1}+2}^{2s} \Delta_{i_{s-m}}\,.
\end{aligned}
\end{equation}
Notice the second line already has the right functional dependence and the right number of sum terms, except for the fact that upper limits in the sums are off. In particular, the last index is not allowed to reach the maximal value $i_{s+1-m}=2(s+1)$. Consider the contribution from $\Delta_{2(s+1)}$ in \eqref{eq:rec-step2}
\begin{equation}
  \Delta_{2(s+1)}\,\omega_j \sum_{m=0}^{s+1} (-1)^m \omega_j^{2m} \sum_{i_1=1}^{2(m+1)} \Delta_{i_1} \dots \sum_{i_{s-m}=i_{s-m-1}+2}^{2s} \Delta_{i_{s-m}}\,.
\end{equation}
In our conventions, the term $m=s+1$ vanishes, be definition. The remaining terms match $\Delta_{2(s+1)}\,Q_{s,j}$ above. The only remaining question is whether the terms in $Q_{s+1,j}$ not including $\Delta_{2(s+1)}$
\begin{equation}
  \omega_j \sum_{m=0}^{s+1} (-1)^m\,\omega_j^{2m}\,\sum_{i_1=1}^{2(m+1)} \Delta_{i_1} \sum_{i_2=i_1+2}^{2(m+2)} \Delta_{i_2}\dots \sum_{i_{s+1-m}=i_{s-m}+2}^{2s+1} \Delta_{i_{s+1-m}}
\end{equation}
equal $\omega_j\,P_{s+1,j}$. Since the upper limit in $i_{s+1-m}$ was reduced to $2s+1$, the remaining upper limits should also be decreased by one, finishing the proof.


\bibliographystyle{utphys}
\bibliography{OpG-2}{}

\end{document}